\documentclass[prx,aps,epsf,twocolumn,showpacs,superscriptaddress,tightenlines]{revtex4-1}
\usepackage{dcolumn}
\usepackage{bm}
\usepackage{epsfig}
\usepackage{latexsym}
\usepackage{amsmath}
\usepackage{amssymb}
\usepackage{color}
\usepackage{array}
\usepackage{bbm}
\usepackage{hyperref}
\usepackage{color}
\usepackage{array}
\usepackage{cancel}
\usepackage[normalem]{ulem}
\usepackage{slashed}

\newcommand{\<}{\langle}

\renewcommand{\>}{\rangle}
\renewcommand{\(}{\left(}
\renewcommand{\)}{\right)}
\renewcommand{\[}{\left[}
\renewcommand{\]}{\right]}

\newcommand{\Ket}[1]{\left|#1  \right>}
\newcommand{\Bra}[1]{\left<#1  \right|}

\newcommand{\tr}{{\mathrm{tr}}}

\begin{document}
\title{Entanglement Transitions from Holographic Random Tensor Networks}

\author{Romain Vasseur}
\affiliation{Department of Physics, University of Massachusetts, Amherst, MA 01003, USA}

\author{Andrew C. Potter}
\affiliation{Department of Physics, University of Texas at Austin, Austin, TX 78712, USA}

\author{Yi-Zhuang You}
\affiliation{Department of Physics, Harvard University, Cambridge, MA 02138, USA} 
\affiliation{Department of Physics, University of California, San Diego, CA 92093, USA} 

\author{Andreas W. W. Ludwig}

\affiliation{Department of physics, University of California, Santa Barbara, CA 93106, USA}

\date{\today}

\begin{abstract}
We introduce a novel class of phase transitions separating quantum states with different entanglement features. An example of such an ``entanglement phase transition'' is provided by the many-body localization transition in disordered quantum systems, as it separates highly entangled thermal states at weak disorder from many-body localized states with low entanglement at strong disorder. In the spirit of random matrix theory, we describe a simple model for such transitions where a physical quantum many-body system lives at the ``holographic'' boundary of a bulk random tensor network. Using a replica trick approach, we map the calculation of the entanglement properties of the boundary system onto the free energy cost of fluctuating domain walls in a classical statistical mechanics model. This allows us to interpret transitions between volume-law and area-law scaling of entanglement as ordering transitions in this statistical mechanics model. Our approach allows us to get an analytic handle on the field theory of these entanglement transitions.

\end{abstract}

\maketitle

\section{Introduction}

Quantum entanglement plays a crucial role in our understanding of non-equilibrium quantum dynamics and the process of thermalization. This can be traced back to the fact that quantum and thermal eigenstates differ dramatically in their entanglement properties. In particular, whereas quantum ground states of gapped systems with local interactions exhibit area-law scaling~\cite{Srednicki1993,Verstraete:2006qt,1742-5468-2007-08-P08024,Eisert:2010rz} with entanglement entropy of a subsystem proportional to its surface area, typical (highly-excited) eigenstates exhibit volume law behavior, with entanglement of a subsystem scaling extensively with its volume~\cite{Page:1993fv,Foong:1994bf,Sen:1996rw,Rigol:2012wd}.

In contrast to basic intuition that states with more entanglement are ``more quantum", extremely entangled states (such as volume-law entangled states) actually exhibit classical dynamics and correlations. Volume-law entanglement implies that every piece of local degree of freedom is highly entangled with the rest of the system, such that most quantum information is scrambled among the whole system and therefore inaccessible to local observables~\cite{Page:1993fv,Hayden:2007uf,Sekino:2008pv,Brown:2012lt,Lashkari:2013mj}. For such highly-entangled eigenstates, only classical hydrodynamic and thermodynamic properties are accessible, which are effectively described by statistical mechanics. This is the essence of the eigenstate thermalization hypothesis~\cite{PhysRevA.43.2046,PhysRevE.50.888} (ETH) which states that the reduced density matrix of a small subsystem in a typical eigenstate of a many-body quantum system is effectively thermal. This implies that the entanglement entropy of small subsystems coincide with the thermodynamic entropy, and is therefore extensive -- {\it i.e.} satisfies a volume-law scaling.

However, not all isolated quantum many-body systems thermalize. In the presence of strong disorder, many-body localization (MBL) can occur, where the excited eigenstates can violate the ETH and exhibit the area-law entanglement~\cite{BauerNayak}. As entanglement formation is obstructed by the strong disorder, an MBL system does not relax to thermal equilibrium, and local quantum information can be preserved coherently for a very long time. 
Highly excited eigenstates of MBL systems therefore have the same area-law entanglement as quantum ground states, opening the door to quantum coherent phenomena in dense, ``hot'' quantum systems far from thermal equilibrium, including symmetry-breaking order, topological order, or even quantum criticality~\cite{Berkelbach:2010ir,PhysRevLett.110.260601,HuseMBLQuantumOrder,BahriMBLSPT,PhysRevB.89.144201,PhysRevLett.112.217204,PhysRevB.90.174302,PhysRevB.91.140202,CenkeMBLSPT,2015arXiv150600592P}.

MBL leads to an entirely new class of dynamical phase transition~\cite{FleishmanAnderson,Gornyi,BAA,PhysRevB.75.155111,PalHuse}  between thermalizing systems, whose long time behavior is described by equilibrium statistical mechanics,  and many-body localized systems, which fail to reach thermal equilibrium even at very long times. While conventional phase transitions separate phases with similar entanglement properties -- with area-law scaling for quantum phase transitions and volume-law scaling for thermal phase transitions, the MBL transition is special in that it is a transition of the entanglement properties of many-body eigenstates, across which the entanglement scaling changes dramatically from the area-law (in the MBL phase) to the volume-law (in the ETH phase). It is also a transition at which the quantum mechanics description is taken over by the (classical) statistical mechanics description, {\it i.e.} \,a transition between quantum and classical phases, which lies beyond the traditional framework of phase transitions. Most of our current understanding of its universal properties relies on small scale numerics~\cite{PalHuse,Luitz,PhysRevLett.113.107204,PhysRevX.5.041047,PhysRevX.7.021013}  and phenomenological renormalization group approaches~\cite{VHA,PVPtransition,PhysRevB.93.224201,2017arXiv170104827D,2017arXiv170609338T,2017arXiv171109880T, 2018arXiv180704285G}. More broadly, many fundamental questions remain: how does the singular entanglement rearrangement occur at the transition? Is the MBL transition the only example of such entanglement transition? Are there other examples of dynamical -- neither quantum nor classical -- transitions that separate quantum states with entanglement properties intermediate between MBL and thermal systems (say with power-law scaling)? 

To address these questions, we propose in this paper to adopt a ``holographic description'' of the entanglement structure of quantum states combined with ideas inspired from random matrix theory. The holographic duality was originally proposed~\cite{Witten:1998ty,Gubser:1998fe,Maldacena1999} as a duality between a quantum field theory and a gravitational theory in one higher dimension. More generally, it is a duality between the entanglement structure of a quantum many-body system on the holographic boundary and the spacetime structure of a gravitational system in the holographic bulk~\cite{Verlinde:2000pn,Ryu:2006fj,Evenbly2011,PhysRevD.86.065007,2012arXiv1209.3304S,Takayanagi:2012vm,Balasubramanian:2013xz,Verstraete:2013do,Maldacena:2013te,Lee:2014qh,Takayanagi:2014an,Leigh:2014rb,Vidal:2014oz,Preskill:2015ja,Takayanagi:2015so,Molina-Vilaplana:2015lr,Takayanagi:2015pd,Czech:2015ig,Bao:2015hf,Lee:2015aa}. The idea is manifested in the Ryu-Takayanagi formula~\cite{Ryu:2006fj} which relates the entanglement entropy of a boundary region to the area of the minimal surface in the bulk that is homologous to the same region. Using this connection,  area-law entanglement corresponds to a shallow holographic bulk with the infrared region capped off, whereas volume-law entanglement typically arises when there is a black hole in the holographic bulk~\cite{Maldacena:2013te,Hayden2016}. This suggests that  entanglement transitions (like the MBL transition) in the quantum many-body system may be viewed as transitions of the holographic bulk geometry in the gravitational dual description~\cite{PhysRevB.93.104205}. In this regard, going one dimension higher into the holographic bulk may provide new insights into the MBL transition. The main goal of this work is to explore one possibility of such holographic description for  entanglement transitions using recent developments in the field of tensor network holography~\cite{Evenbly2011,PhysRevD.86.065007,2012arXiv1209.3304S,Vidal:2014oz,Preskill:2015ja}.

\begin{figure}[t!]
\begin{center}
\includegraphics[width=\linewidth]{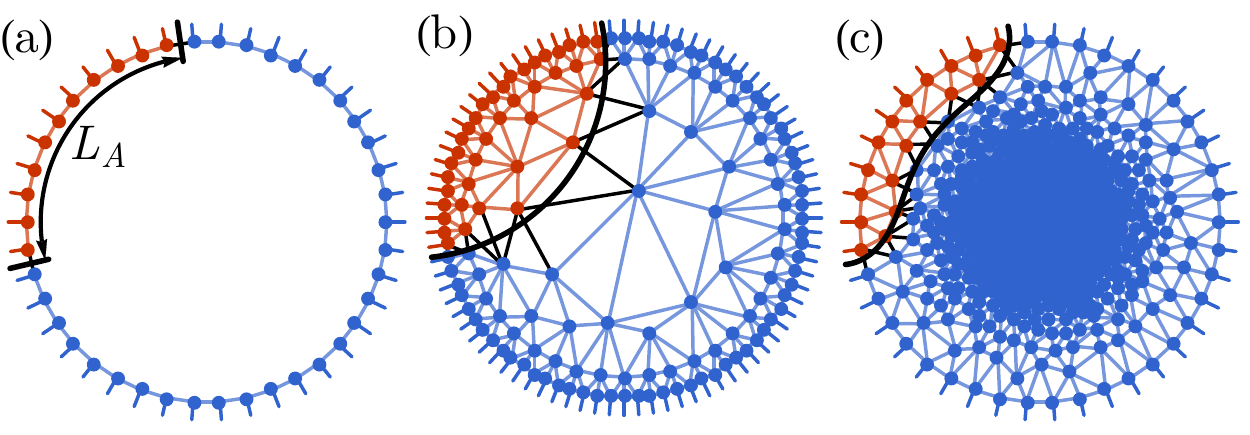}
\caption{Different tensor network geometries correspond to different entanglement scalings for large bond dimension: (a) quasi-linear geometry (MPS-like) and area-law entanglement ($S_A\sim\text{const.}$), (b) hyperbolic geometry and logarithmic entanglement scaling ($S_A\sim\log L_A$), and (c) ``black hole'' geometry and volume-law entanglement ($S_A\sim L_A$).}
\label{fig:intro}
\end{center}
\end{figure}

The tensor network approach is both an efficient mathematical construction to represent quantum many-body states and a powerful tool to capture different entanglement patterns~\cite{doi:10.1080/14789940801912366,1751-8121-42-50-504004,Evenbly2011}. It underlies the success of many variational tensor-network state algorithms for computing ground-state properties, such as the density matrix renormalization group (DMRG) in one dimension~\cite{White:1992lo,Schollwock:2011ud}. Tensor network constructions have also been employed to model the holographic duality. A generic tensor network description for a one dimensional quantum state should be defined in a two dimensional holographic bulk, as shown in Fig.~\ref{fig:intro}, where each dot represents a tensor and the tensor indices are contracted on each link that connects the tensors. The physical degrees of freedom live on the boundary of the network (represented by  the un-contracted dangling legs). The tensor contractions result in a complex number that describes the many-body wave function for the physical system. The bulk tensors are introduced as hidden variables, which are necessary to capture the complicated multi-body entanglement in the many-body state. The tensor network geometry depends on the entanglement structure of the boundary quantum state: for area-law entangled states, one could use linear (or quasi-linear) graphs (with no or little bulk layers) like Fig.~\ref{fig:intro}(a), reducing the tensor network to matrix product states (MPS) routinely used in DMRG simulations~\cite{Schollwock:2011ud}; whereas critical states with logarithmic entanglement require a more involved tree-like hyperbolic bulk structure called MERA~\cite{PhysRevLett.99.220405,PhysRevLett.101.110501}, similar to Fig.~\ref{fig:intro}(b). There are procedures to efficiently construct explicit tensor network representations for MBL (or marginal MBL) states~\cite{PhysRevB.92.024201,PhysRevB.95.035116,PhysRevB.94.041116,PhysRevB.93.104205,PhysRevLett.116.247204,PhysRevLett.118.017201,PhysRevX.7.021018}. However, for volume-law entangled states on the ETH side, the holographic tensor network would involve a densely connected subgraph (or even just a big random tensor) as in Fig.~\ref{fig:intro}(c) to represent a black hole state. Currently, there is no numerically efficient method to manipulate tensor contractions on such networks. This renders the exact computation of the ETH state unaffordable for large sized systems, which also hinders the study of MBL transition that is adjacent to the ETH phase.

On the other hand, extracting statistics of properties such as entanglement or hydrodynamics over the ensemble of thermal states may not require having access to the full structure of exact many-body wave functions. Such an idea has its conceptual roots in the development of the quantum chaos theory in terms of random matrices~\cite{RevModPhys.53.385}, where thermal states were simply modeled by random states (or random tensors in the tensor network language), and this philosophy has been exploited in several recent works to study entanglement dynamics and hydrodynamics in random quantum circuit evolutions~\cite{PhysRevX.7.031016,2017arXiv170510364N,2017arXiv170508975N,2017arXiv171009835K,2017arXiv170508910V,2017arXiv171009827R,2017arXiv171206836C,2018arXiv180409737Z}.

In this work, we explore the structure of entanglement, in a refined version of the random matrix ansatz, where instead of a big fully connected bulk tensor, we consider a structured bulk network of locally connected tensors, in which each tensor is randomly drawn from a uniform distribution, producing an ensemble of random tensor network (RTN) states, as proposed by~\cite{Hayden2016,Yang:2016fu,Qi:2017qf}. We will develop a new approach to these RTN states that enables a rare analytic handle on their entanglement properties. Further, in analogy to the black hole no hair theorem~\cite{Misner:1973sb,Israel:1967kr,Israel:1968sv,Carter:1971sw}, that the area of the black hole is proportional to its entropy and is independent from details of its initial state, we focus on the statistics of entanglement entropies rather than detailed structure of the entanglement spectrum and eigenstates (Schmidt states). The RTN wavefunctions we consider are exactly the type of ``bald" networks, average over random tensors effectively removes all features in the many-body state that are not invariant under local unitaries, leaving only the entanglement features encoded in the network structure.

Given a quantum state $\Ket{\psi}$ of the one dimensional boundary system with density matrix $\rho=\Ket{\psi} \Bra{\psi}$  obtained from such a random tensor network defined on a planar graph $G$, we are interested in the entanglement entropy $S_A = - {\rm tr} \rho_A \log \rho_A$ for the reduced density matrix  $\rho_A = {\rm tr}_{\overline{A}} \rho /  {\rm tr} \rho$ with $A$ a given subregion of the physical system. This quantity was computed in the limit of strictly infinite bond dimension in Ref.~\cite{Hayden2016}, corresponding to maximally entangled states (deep in the thermal phase).
Building on these results, we will introduce a generalized replica trick to move away from this restrictive limit, and explore the physics of random tensor networks at arbitrary bond-dimension to explore both volume and area-law states. In the holographic language, this corresponds to incorporating quantum effects beyond the classical gravity description of the bulk geometry. This allows us to access a quantum regime where the Ryu-Takayanagi formula breaks down.  

 In particular, we show that the  entanglement entropy (and all R\'enyi entropies ${\rm tr} \rho_A^m$) averaged over all possible random tensor realizations can be computed for {\em any bond dimension} by using a generalized replica trick and mapping the entanglement calculation onto a classical spin model defined on the graph $G$. In that language, we can interpret the quantum entanglement entropy of the boundary physical system as the free-energy cost of a boundary domain in the classical spin model. By tuning the bond dimension of the random tensor networks, this construction allows us to identify an entanglement transition of  the quantum wave function as the paramagnetic-to-ferromagnetic transition in the classical spin model. This gives us an analytic handle on the universal properties and the scaling of entanglement at the critical point, and provides the first field theory formulation of such entanglement transitions. By considering the graph $G$ to be random, we uncover a new dynamical transition separating phases with area-law and power-law scaling of entanglement. The universal properties of these entanglement transitions can be described using a certain conformal field theory (CFT) with central charge $c=0$ coupled to two-dimensional quantum gravity, leading to exact relations for the critical exponents.

The remainder of this paper is organized as follows: in section~\ref{secRTN}, we introduce our random tensor network model, and show how the entanglement features of that network can be computed using a classical statistical mechanics model 
and  a replica trick approach. We discuss the limiting cases of large and small bond dimension in section~\ref{secLimitingCases}, and we then propose an analytic continuation of the statistical mechanics model that allows us to access the universal features of the entanglement transition between these two regimes (section~\ref{secUniversalFeatures}). We also describe how the graph on which the tensor network is defined can be made random, which translates in the field theory language as a coupling of the statistical model to two-dimensional quantum gravity (section~\ref{SecRandomGravity}). Finally, section~\ref{SecDiscussion} contains a discussion of the applicability of our results to the MBL/ETH transition, as well as possible extensions of this work.

\section{Random Tensor Networks and Entanglement}

\label{secRTN}

\subsection{Random projected entangled pair states (PEPS)}
To establish notation, we briefly review the projected entangled pair states (PEPS)~\cite{2004cond.mat..7066V} construction of a random tensor network~\cite{Hayden2016}. Consider a quantum system (henceforth called the physical system) whose Hilbert space is a product of on-site Hilbert spaces: $\mathcal{H}_\text{phy} = \otimes_{i\in \text{phy}}\mathcal{H}_i$. For convenience, we consider each $\mathcal{H}_i$ to be of fixed dimension, $D_p$ (e.g. $D_p=2$ for a spin-1/2 Hilbert space).

\begin{figure}[tb]
\begin{center}
\includegraphics[width=\linewidth]{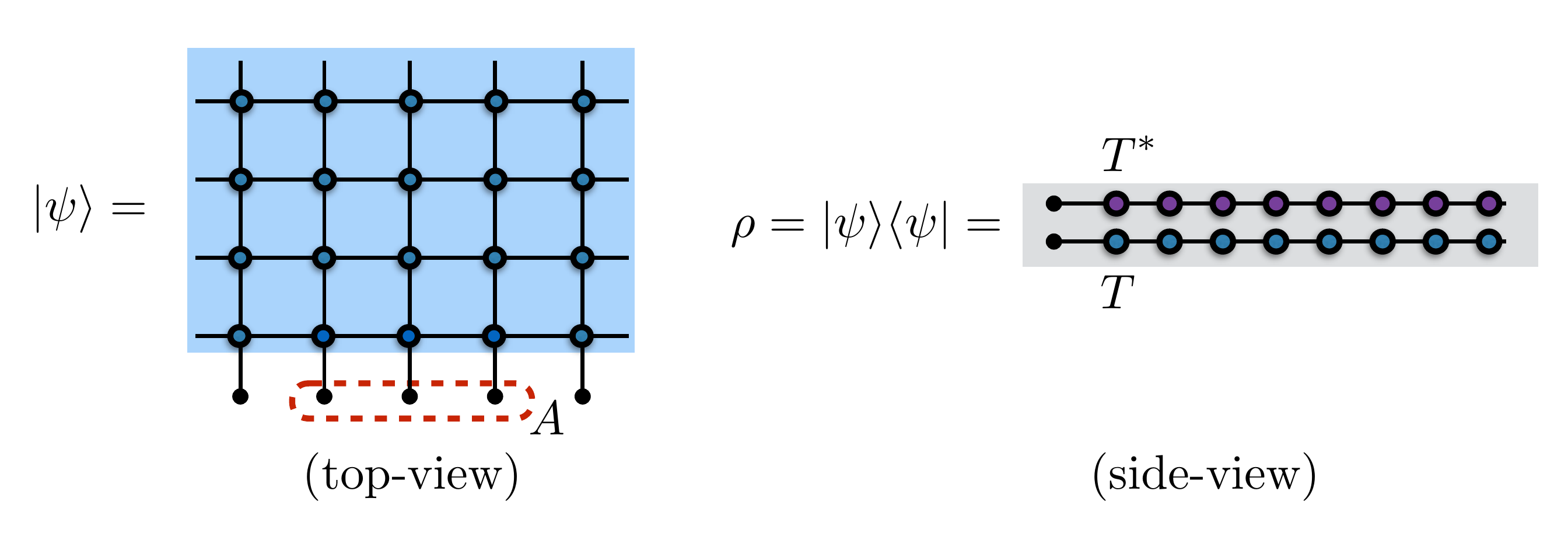}
\vspace{-.1in}
\end{center}
\caption{
{\bf Random tensor network state -- } Schematic of wave-function $|\psi\>$ (left-panel, top-view) and density matrix $\rho = |\psi\>\<\psi|$ (right-panel, shown in side-view, corresponding to a fixed column cross-section of the left-panel). Colored circles represent tensors, with blue circles ($T$) for the ``ket" state's tensors and purple dots for the ``bra" state's tensors ($T^*$). The black lines connecting tensors represent PEPS contractions of internal states, and dotted dangling lines correspond do physical degrees of freedom. }
\label{fig:rtn}
\end{figure}

A PEPS wave-function can be constructed by augmenting the physical space by auxiliary quantum degrees of freedom (DOF) defined on a graph (network) $G=(V,E)$, with vertices (sites) $V$ and edges (bonds) $E$. We will refer to the auxiliary DOF as the ``bulk", marked out by the blue region in  Fig.~\ref{fig:rtn}.  
 The physical DOF are then assigned to dangling ends on the boundary of the graph $G$, represented by the small black dots in 
 Fig.~\ref{fig:rtn}.
 For each pair $(v,e)$ of site $v\in V$ and its adjacent bond $e\in E$, we define an associated Hilbert space $\mathcal{H}_{ve}$ of dimension $D_e$, which can be spanned by a set of basis states $|\mu_{ve}\>$ labeled by $\mu_{ve}=1,2,\cdots,D_e$. In the tensor network language, $D_e$ is the bond dimension specific to each bond $e$.

On each bulk site $v\in V$, i.e.~each small black circle in 
 Fig.~\ref{fig:rtn},
we define a state (the site state)
\begin{equation}
|T_v\>=\sum_{\{\mu_{ve_i}\}}(T_v)_{\mu_{ve_1}\cdots\mu_{ve_z}}|\mu_{ve_1}\>\cdots |\mu_{ve_z}\>,
\end{equation}
where $e_i$ ($i=1,\cdots,z$) in $\mu_{ve_i}$ denote the bonds emanating from the site $v$. The number of adjacent bonds of a site (the degree of the vertex) is denoted as the coordination number $z$ of that site. Each site state $|T_v\>$ is specified by a set of coefficients $(T_v)_{\mu_{ve_1}\cdots\mu_{ve_z}}$, which can be further arranged into a tensor $T_v$. Therefore each bulk site $v$ represents a tensor $T_v$ in the tensor network. A special subset of vertices on the boundary of the graph (small black dots in Fig.~\ref{fig:rtn}(a)) are taken to be the physical DOF, on which the physical Hilbert spaces $\mathcal{H}_i$ are defined. These boundary vertices are dangling. No site states are defined on them. 

On each bond $e\in E$, i.e.~each link in Fig.~\ref{fig:rtn}(a), we define an entangled pair state $|I_e\>$ (the bond state),
\begin{equation}\label{eq: Ie state}
|I_e\>=\sum_{\mu_{ve},\mu_{v'e}=1}^{D_e}\lambda_{\mu_{ve}\mu_{v'e}}|\mu_{ve}\>
|\mu_{v'e}\>,\end{equation}
where $v$ and $v'$  denote the two sites linked by the bond $e$. Let us treat the coefficients $\lambda_{\mu_{ve}\mu_{v'e}}$ as a matrix and define the $n$th Renyi mutual information across the bond $e$ as
\begin{equation}\label{eq: Ie def}
I_e^{(n)}=\frac{2}{1-n}\log\mathrm{Tr}(\lambda\lambda^{\dagger})^{n}.
\end{equation}
The entangled pair state $|I_e\>$ across the bond $e$ is then characterized by the set of 
values of the bond mutual information $I_e^{(n)}$, which can change continuously from $0$ to $2\log D_e$ in general. The maximal mutual information $I_e^{(n)}=2\log D_e$ can be achieved if the state $|I_e\>$ is maximally entangled, e.g.~$\lambda_{\mu_{ve}\mu_{v'e}}=D_e^{-1/2}\delta_{\mu_{ve}\mu_{v'e}}$.  However considering more general (and less entangled) states $|I_e\>$ allows us to treat $I_e^{(n)}$ as continuously tunable parameters (rather than discretely tuned by $D_e$), which can be used to control the entanglement properties of the PEPS wave-function and to drive the entanglement transition.
Finally, we will always assume that the boundary vertices are connected to the bulk vertices by maximally entangled states with fixed bond mutual information $2\log D_p$.

With $|T_v\>$ and $|I_e\>$ defined above, the final step is to project the bulk states $|T\>=\otimes_{v\in V}|T_v\>$ onto the entangled pair states on the bonds of the graph, 
\begin{equation}
|\psi[T]\> =\bigotimes_{v\in V}\bigotimes_{e\in E}\<T_v|I_e\>.
\label{eqWavefunction}
\end{equation}
The remaining boundary vertices that are not touched by the projection will provide the physical degrees of freedom, on which the PEPS wave-function $|\psi[T]\>$ is supported.
Equivalently, we may work with the density matrix:
\begin{align}
\rho[T] = |\psi[T]\>\<\psi[T]|.
\end{align} 

As a mock-up of wave-functions for complicated, disordered, and interacting quantum systems, in this work, we will study the entanglement properties of \emph{random PEPS states}, or \emph{random tensor network} (RTN) states, obtained by drawing the tensor $T_v$ for each site $v$ independently from a featureless Gaussian distribution characterized by zero mean, $\overline{T_v}=0$, and variance: 
\begin{align}\overline{ (T^*_v)_{\mu_1\dots \mu_{z}}(T_{v})_{\nu_1\dots \nu_{z}} } = \delta_{\mu_1,\nu_1}\dots \delta_{\mu_{z},\nu_{z}}.
\end{align} 
Importantly, we note that the random PEPS states, as defined, are unnormalized. Hence, to properly compute
disorder averaged quantities we will need to explicitly normalize these wave-functions before computing disorder averages. This normalization issue will be a key technical challenge in what follows, and its resolution will present our primary departure from the work of Hayden et al~\cite{Hayden2016}, which allows us to explore phase transitions in the entanglement structure of PEPS wave-functions.

\subsection{Entanglement of Random PEPs}

We would like to compute the entanglement of a region $A$ of the physical system in the random PEPs state $|\psi[T]\>$, measured in terms of the tensor-averaged Renyi entropies
\begin{align}
S^{(n)}_A[T] = \frac{1}{1-n}\overline{\log\(\frac{\text{tr}\rho_A^n}{\(\text{tr}\rho\)^n}\)},
\end{align}
where $\overline{(\cdots)}$ refers to averaging over random tensors; we define here
$\rho_A\equiv \text{tr}_{\bar A} \rho$, Fig.~\ref{fig:rtn}(a) and we have explicitly divided by $\text{tr}\rho$ to ensure the normalization of the random PEPs density matrix for each realization of disorder.

In Ref.~\cite{Hayden2016}, the tensor-average was performed in the limit of large bond dimension $D_e\rightarrow \infty$, where the averages of
the logarithms
 could be replaced by
logarithms of the averages,
 i.e.:
\begin{align}
\lim_{D_e\rightarrow \infty} S^{(n)}_A[T] = \frac{1}{1-n}\[\log \overline{\text{tr}\rho_A^n}-\log\overline{\text{tr}\rho^{\otimes n}}\].
\end{align}
However,
exchanging of the order 
of tensor-averaging and taking the logarithm works only in the limit of infinite bond dimension. In fact, Hayden et al. estimate that these expressions are valid only if the bond dimension diverges as a power $p$ in the physical system size, $L$: $D_e\gg L^p$~\cite{Hayden2016}, such that these expressions do not extend sensibly to thermodynamically large systems. 

We would like to extend these results to the physically reasonable regime of finite bond dimension, and in particular to examine the limits of validity in the conjectured correspondence between random tensor networks~\cite{Hayden2016} and the geometric Ryu-Takayanagi formula for entanglement~\cite{Ryu:2006fj}, and to access a phase transition in the entanglement structure of these RTNs between area and volume law entanglement scaling. 
%
To this end, we introduce a replica trick to properly average over random tensors for \emph{arbitrary} bond-dimension. Namely, we consider taking $m$ fictitious copies, or ``replicas" of the system. 
The  Renyi entropies for a generic (unnormalized) random PEPs state can then  be written as:
\begin{align}
S^{(n)}_A = \(\frac{1}{1-n}\)\lim_{m\rightarrow 0}
\frac{1}{m} \[\overline{\(\text{tr}\rho_A^n\)^m}-\overline{\(\text{tr}\rho^n\)^m}\].
\label{eq:replica}
\end{align}
Crucially, this formulation works for arbitrary random PEPs states, without any assumption on geometry, dimensionality, tensor structure, or bond dimension. Using this replica formulation, we will confirm that the infinite bond dimension results describe the physical properties of RTNs above a critical bond dimension $D_c$, but qualitatively break down for $D_e< D_c$, leading to an eigenstate phase transition and a breakdown of the Ryu-Takayanagi formula~\cite{Ryu:2006fj} and the corresponding connections to dual theories of gravity.


\subsection{Mapping to spin model -- Result}
To utilize this replica formulation of the entanglement of RTNs, we will generalize the observation of Hayden et al.~\cite{Hayden2016}, that powers of the reduced density matrix for a RTN can be mapped to a classical spin model. In this section, we review these results, which establish the mapping.
In particular, we write
\begin{align}
\overline{(\text{tr} \rho^n)^m}=\overline{(\text{tr} \rho)^{nm}}= \mathcal{Z}_0^{(nm)},
\end{align}
where $(\text{tr} \rho^n)^m$ and $(\text{tr} \rho)^{nm}$ are equivalent as long as $\rho$ is pure, and
\begin{align}
\mathcal{Z}_0^{(nm)} = \sum_{g_v\in S_{nm}} e^{-\sum_{\<vv'\>} H_{\<vv'\>}\(g_v^{-1}g_{v'}\)},
\label{eqSpinModel}
\end{align}
is the partition function for a spin model with spins on the sites of the tensor network taking values in the permutation group
$S_Q$ permuting
$Q\equiv n\times m$ elements: $g_v\in S_{nm}$, $v\in V$. Here $H_e(g)$ is a bond-specific
function which depends only on the conjugacy class of the group element 
 $g\in S_{nm}$ (a ``class function''), and which reads explicitly
\begin{equation}\label{eq:He in Ie}
H_e(g)=\sum_{\alpha=1}^{C(g)}\frac{l_g^\alpha-1}{2}I_e^{(l_g^\alpha)}.
\end{equation}
Here  $l_g=(l_g^1,l_g^2,\cdots)$  denote the cycle type of the permutation $g$, where $l_g^\alpha$ is the length of the $\alpha$th cycle in $g$ and the total number of cycles is denoted by $C(g)$. If the bond states $|I_e\>$ are taken to be maximally entangled, the class function $H_e(g)$
simplifies to 
\begin{equation}\label{eq:He=-JeC}
H_e(g)=-J_e C(g)+J_e n m,
\end{equation}
where $C(g)$ is the cycle counting function  defined above, and the interaction strength $J_e=\log D_e$ is set by the bond dimension $D_e$. An unimportant constant $J_e nm$ comes from the normalization of the state $|I_e\>$ and can be dropped. Note that if the bond-dimension is inhomogeneous, the effective interactions will have the same inhomogeneity. The interaction $C(g)$ gives a basis independent measure of how far $g=g_v^{-1}g_{v'}$ would be from the identity. This function is maximal for the identity: $C(1) = nm$, and achieves a minimum of $1$ for a full cyclic permutation of all $nm$ elements (Fig.~\ref{fig:perms}).We will mainly focus on the case of maximally entangled $|I_e\>$, where the ``spin interaction'' energy $H_e(g_v^{-1}g_{v'})$ is simply given by the cycle counting function $C(g_v^{-1}g_{v'})$ as in Eq.~\ref{eq:He=-JeC}, such that neighboring $g_v$ and $g_{v'}$ are favored to be the same.

\subsection{Mapping to spin model -- Derivation}
We would like to average quantities such as: $\rho^{\otimes Q}[T]$, for integer $Q=nm$, over the random tensors. We will take the tensors to be
independent identically distributed ( i.i.d.)  featureless Gaussian 
random variables, whose probability distribution is  characterized by zero mean, $\overline{T_v}=0$, 
 variance,  $\overline{ (T^*_v)_{\mu_1\dots \mu_{z}}(T_{v})_{\nu_1\dots \nu_{z}} } = \delta_{\mu_1,\nu_1}\dots \delta_{\mu_{z},\nu_{z}}$, and higher moments given by Wick decomposition.

\begin{figure}[tb]
\begin{center}
\includegraphics[width=0.8\linewidth]{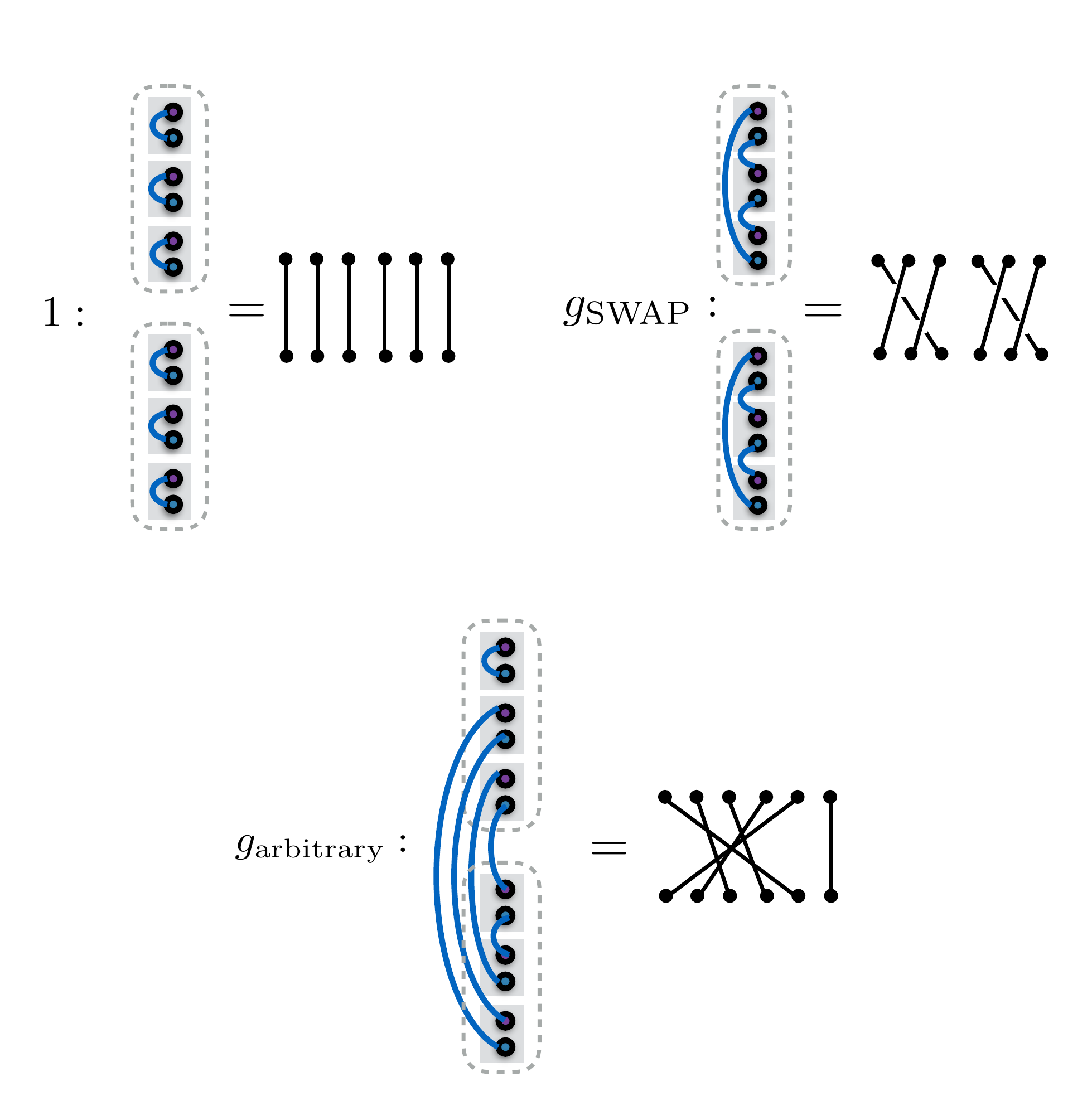}
\vspace{-.1in}
\end{center}
\caption{
{\bf  Examples of Wick contractions and corresponding permutation elements -- } Averaging over disorder results in Wick contractions (blue curved lines) of each tensor of the ``ket" state (blue dot) of each copy of the density matrix, with that of a ``bra" tensor (purple dot) of another replica copy. This defines a permutation element. Three examples are shown for $n=3$ and $m=2$. For clarity, only one tensor site in the network is shown, and the corresponding permutation element is drawn. (Top-left) identity permutation, $1$: each tensor is contracted within the same copy of $\rho$, (top-right) Swap permutation, $g_\text{SWAP}$ permutation: the tensors are combined into $m$ groups of $n$, and for each group, each tensor is contracted with the next tensor in a cyclic sequence, (bottom) arbitrary permutation: a generic contraction and corresponding permutation element.
}
\label{fig:perms}
\end{figure}

Then, to obtain a non-zero contribution to $\overline{\rho^{\otimes Q}}$, we must Wick-contract each tensor for each vertex in $|\psi\>$ with a complex conjugated counterpart in one of the replica copies of $\<\psi|$. The set of possible contractions can be labeled by a permutation group element 
of $Q$ 
copies  for each vertex, $g_v\in S_Q$, such that $\overline{\rho^{\otimes Q}} = \sum_{\{g_v\}} w(\{g_v\})$ for some weight function $w$ (Fig.~\ref{fig:perms}). 

\begin{figure}[tb]
\begin{center}
\includegraphics[width=\linewidth]{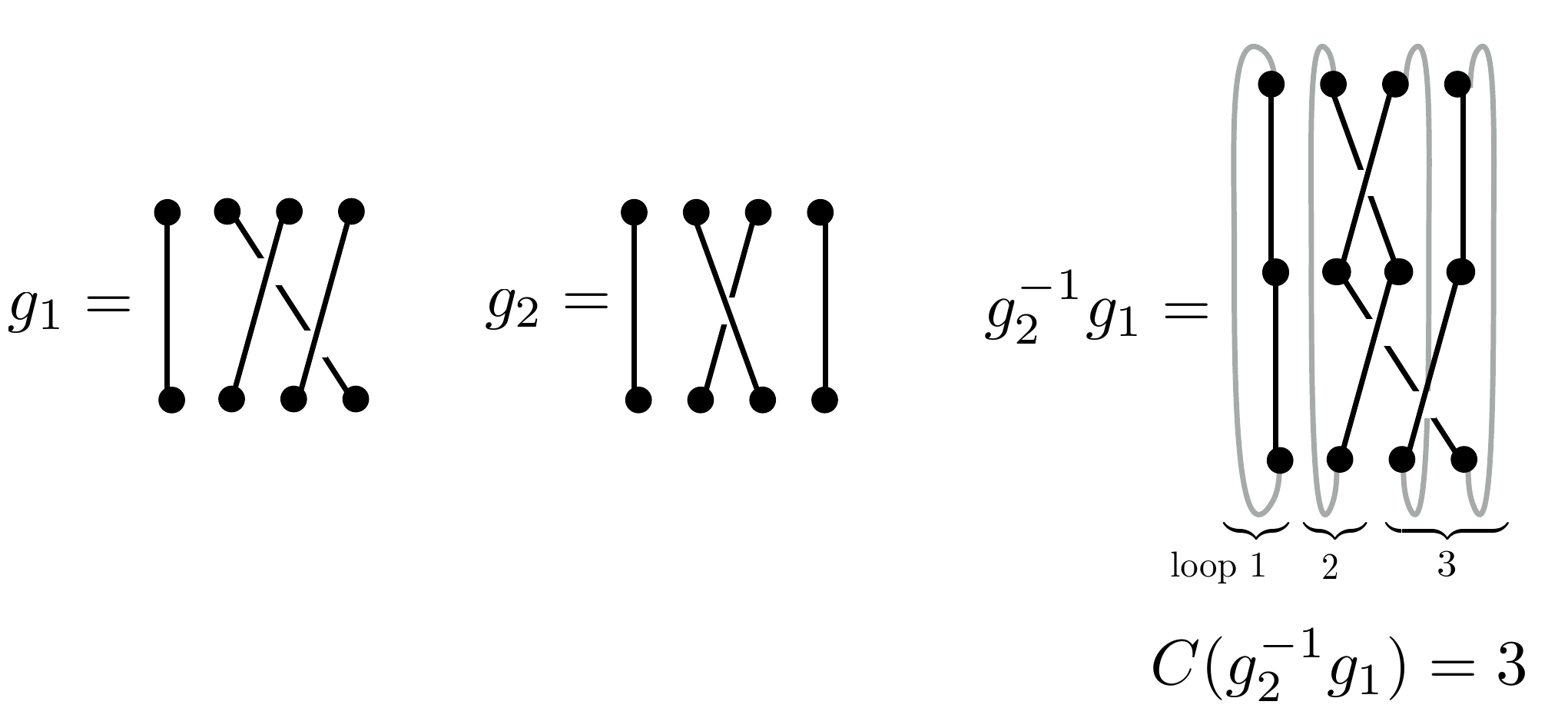}
\vspace{-.1in}
\end{center}
\caption{
{\bf  Permutation character -- } Schematic representation of two permutations, $g_{1,2}$, on $6$ items. In the spin model, the overlap of two neighboring permutation spins is measured by the group character $C(g_1^{-1}g_2)$, which can be computed by concatenating $g_2$ and $g_1^{-1}$ (right panel), connecting the final points directly back to the initial ones (gray lines), and counting the number of independent loops (in this case, $3$). 
}
\label{fig:character}
\end{figure}

{\bf General structure of $w$ -- } Before working out the full expression for $w$, we deduce some general features. First, $w$ is positive definite. Next, since left- or right- multiplication of all $g_i$ by the same permutation element $h\in S_Q$ is simply a re-ordering of the identical factors of $\rho$ in $\rho^{\otimes Q}$, the weight function is invariant under both left- and right- multiplication by $h$. Hence, the symmetry group of the spin model will be $S_Q\times S_Q$. Furthermore, since, in our class of tensor network wavefunctions, the tensors on two vertices are only connected if they share a common bond, the weight should factorize into a product of pairwise weights, which we are free to write in the Boltzmann form: $w(\{g_v\})\equiv e^{-\sum_{\<vv'\>}H_{vv'}(g_v^{-1}g_{v'})}$. From these general considerations we immediately see that the $\overline{\rho^{\otimes Q}}$  takes the form of the partition function of a classical spin model with spins taking values in
the permutation group  $S_Q$ of $Q=n m$ elements on each vertex of the tensor network.

{\bf Explicit form of $w$ -- } Given the general structure above, we can work out the explicit form of $w$. Because $w$ can be factorized
as a product over bonds  as follows,
\begin{equation}
w(\{g_v\})=\prod_{\<vv'\>} w_{\<vv'\>}(g_v^{-1}g_{v'}),
\end{equation}
we only need to focus on one particular bond. Let us consider the bond $e=\<12\>$ linking the sites $1$ and $2$ in the network. We will first consider the bond state $|I_e\>$ to be maximally entangled. Let us label the auxiliary Hilbert spaces for replica $a$ on bond $e$ by a basis of states $|\mu_1^a,\mu_2^a\>$, where $\mu_1\in \{1\dots D_e\}$, and $a=1\dots Q$ is a replica index. 

Let us work out the weight of a specific set of Wick contractions for averaging the random tensors. Contracting a ``bra" tensor for site $1$ from replica $a$ with a ``ket" tensor in replica $b$ forces the indices $\mu^a_1 = \mu^b_1$ for all bonds emanating from site $1$. Moreover, performing the projection to the maximally entangled pair on a bond forces the indices for the tensors on both ends of that bond to coincide. Then, all possible configurations are summed over, and the weight is given by the number of states satisfying these two constraints. Due to the
symmetry under left- and right- multiplication,
$g_v \to g_v h_R$,
$g_{v'} \to g_{v'} h_R$without loss of generality, we may take $g_1=1$, and consider arbitrary $g_2$. We can write $g_2$ in cycle-notation, 
e.g. 
when $Q=5$, the permutation $(134)(25)$ cyclically permutes $1\rightarrow 3\rightarrow 4\rightarrow 1$, and $2\leftrightarrow 5$. The joint effect of contractions for the tensors $T_1$ and $T_2$ at the two sites $1$ and $2$,  along with the entangled pair projection at the bond $e=(12)$ linking these two  sites, constrains the indices $\mu^a$ to coincide for all auxiliary DOF within each cycle. Hence, the number of states for this contraction is equal to $D_e^{C(g_2)}$, where $C(g)$ is the number of distinct cycles in $g$ (e.g. $C(g) = 2$ for the above example, $g=(134)(25)$). For general $g_1$ and $g_2$, this shows that the weight is
\begin{equation}
w_e(g_1^{-1}g_2)\propto D_e^{C(g_1^{-1}g_2)} = e^{J_e C(g_1^{-1}g_2)},
\end{equation}
with $J_e=\log D_e$ as claimed previously.

If we further consider the bond state to be a generic entangled pair state $|I_e\>$  from
 Eq.~\ref{eq: Ie state} above, the weight $w_e$ on the bond $e$ will be generalized to
\begin{equation}
w_e(g)=\prod_{\alpha=1}^{C(g)}\tr (\lambda\lambda^\dagger)^{l_g^\alpha}=\prod_{\alpha=1}^{C(g)}\exp\Big(\frac{1-l_g^\alpha}{2}I_e^{(l_g^\alpha)}\Big),
\end{equation}
in terms of  the values of the bond mutual information $I_e^{(n)}$ defined in Eq.~\ref{eq: Ie def}. So the energy function $H_e$ can be read out from the Boltzmann form $w_{\<12\>}=e^{- H_{\<12\>}(g_1^{-1}g_2)}$, and the result was given by Eq.~\ref{eq:He in Ie}. In general, $H_e$ is continuously tunable by the bond mutual information~\cite{2018PhRvB..97d5153Y}, which provides us the flexibility to drive the RTN state through the entanglement transition.

\noindent{\bf Boundary conditions:}
So far, we have been considering the mapping of the bulk of the tensor network to a classical permutation-spin model. It is also important to properly fix the boundary conditions to perform the appropriate contraction of the $n\times m$ replicas of $\rho$. Consider computing the entanglement entropy of a physical (boundary) region $A$. To apply Eq.~\ref{eq:replica}, we can first consider $nm$ copies of the system's density matrix: $\otimes^{nm}\rho$. Let us group these $nm$ factors of $\rho$, into $m$ groups of $n$, and order the factors in each group from $1\dots n$. Each copy of the density matrix, reads: $\rho = |\psi\>\<\psi|$, where $|\psi\>$ is the PEPS wave-function of the network, i.e. contains a vector (``ket", $|\psi\>$) with tensors $T$, and a dual vector (``bra", $\<\psi|$) with conjugated tensors $T^*$.

\begin{figure}[tb]
\begin{center}
\includegraphics[width=\linewidth]{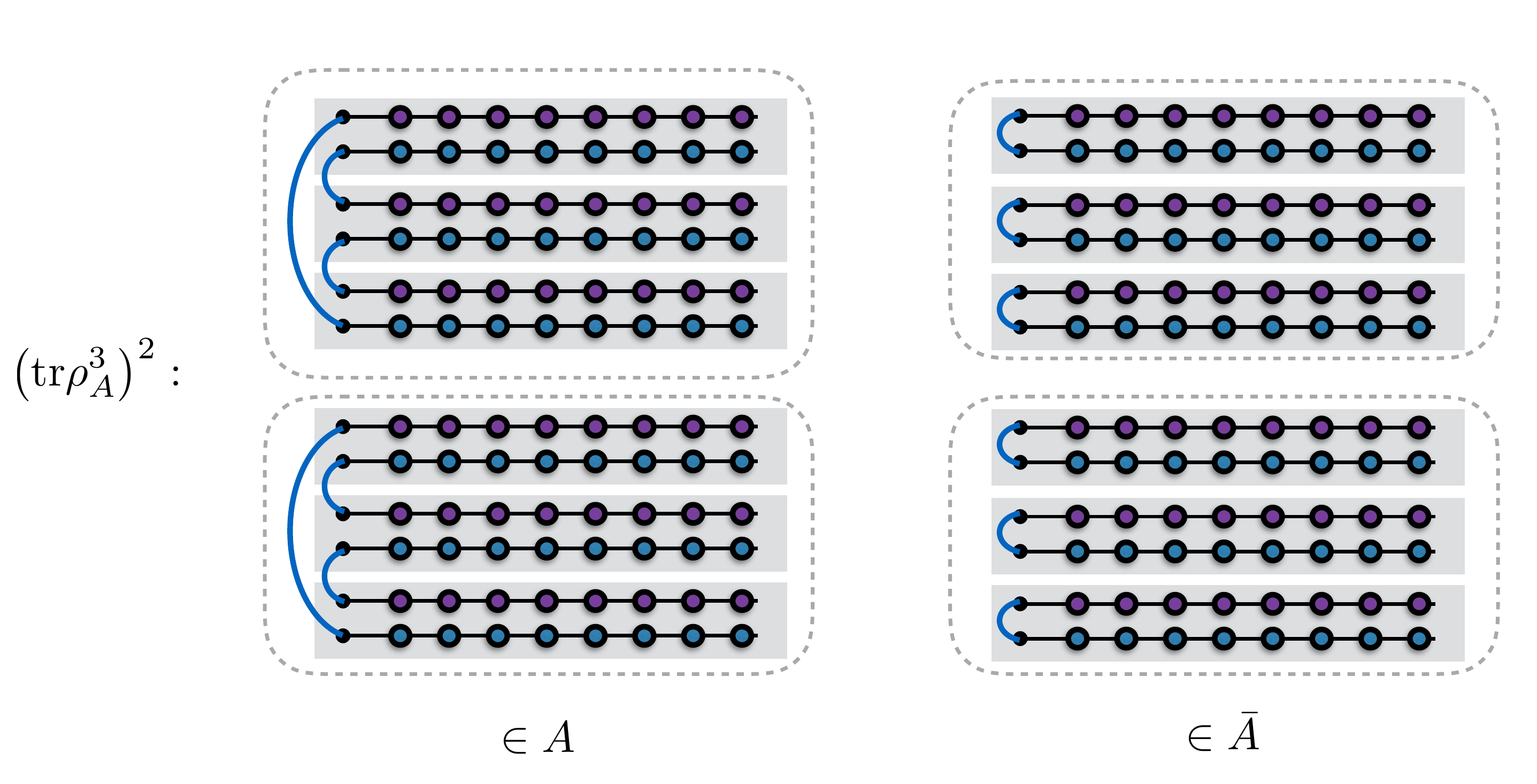}
\vspace{-.1in}
\end{center}
\caption{
{\bf  Boundary conditions -- } Schematic representation of boundary conditions in the entanglement region $A$ (left-panel) and in its complement, $\bar{A}$, (right-panel) for the quantity $\(\text{tr}\rho_A^n\)^m$ the case of $n=3$, $m=2$.}
\label{fig:bcs}
\end{figure}

To compute the second term, $\mathcal{Z}^{(nm)}_0= \overline{(\tr \rho^n)^m}$, in Eq.~\ref{eq:replica},
one contracts within each group of $n$ copies of $\rho$
 all of the physical (boundary) legs of the vector (``ket") of the $j^\text{th}$ copy of $\rho$ to those of the dual vector (`` bra") in the $\((j+1)~\text{mod}~n\)^{\text{th}}$ copy (see Fig.~\ref{fig:bcs}). In contrast, to compute the
first term $\mathcal{Z}^{(nm)}_A=\overline{\(\text{tr}\rho_A^n\)^m}$
involving the reduced density matrix of $A$, one should contract the physical legs inside $A$ as for $\mathcal{Z}^{(nm)}_0$, but, to trace out $\bar{A}$, one should connect the physical legs of the ``ket"-tensors in the $j^\text{th}$ copy to those of the ``bra"-tensors in the same copy (Fig.~\ref{fig:bcs}). 

In other words, in
 the complement
$\bar{A}$
of the entanglement region,
the boundary tensors
 in $\mathcal{Z}^{(nm)}_A$
 are contracted within each copy of $\rho$, without permutation, corresponding to a trivial (identity) permutation between the ``bra" and ``ket" factors of $\rho^{\otimes nm}$. In contrast, the pattern of contractions 
 in $\mathcal{Z}^{(nm)}_A$ 
 inside the entanglement region $A$
corresponds to a non-trivial permutation element that cyclically permutes the ``bra" and ``ket" factors within each of the $m$ factors of $\rho^{\otimes n}$. We 
denote this permutation (as it is convention) by $g_\text{SWAP}\in S_{nm}$, which will play a special role in what follows (Fig.~\ref{fig:perms}). On the other hand, the boundary tensors in $\mathcal{Z}^{(nm)}_0$ are contracted with a fixed (arbitrary) permutation everywhere at the boundary, which can be chosen by symmetry to be the identity since $\rho$ is pure, corresponding to writing $\mathcal{Z}^{(nm)}_0=\overline{(\text{tr} \rho)^{nm}}$. 

Once the boundary contractions have been performed, one can average over the random tensors, resulting in a Wick contraction of each ``ket" tensor $T$ in one copy, $j\in 1,\dots, nm$, with the ``bra" tensor, $T^*$ of another copy $g(j)$ as described above, resulting in a classical spin model with spins defined by the permutation $g$ for each tensor, as described above. 

The fixed boundary conditions for the physical tensor legs  give a larger weight (in the random-tensor average), when the permutation  group element corresponding to the Wick-contraction of the boundary tensors in the tensor-averaging  matches the permutation group  element characterizing the contractions in $(\text{tr}{\rho_A^n})^m$ (either $g_v=g_\text{SWAP}$ for $v\in A$ or $g_v=1$ in $v\in\bar{A}$). This effect can be captured by a boundary ``field" that breaks the permutation symmetry, described by a Hamiltonian:
\begin{align}
\label{LABELEqBoundaryMagneticFields}
H_\text{bdry}=-\sum_{v\in A} h_v C(g_\text{SWAP}^{-1}g_v)-\sum_{v\in \bar{A}} h_v C(g_v),
\end{align}
where $h_v= \log D_p$ is the log of the physical onsite Hilbert space dimension.

\subsection{Physical Picture} 
In the spin-model, the different boundary fields in $A$ and $\bar{A}$, effectively ``twist" the boundary spins from
$g_\text{SWAP}$ (in $A$) to
 $1$ (in $\bar{A}$). If the boundary spins follow this field then, this twist introduces an extra spin domain wall terminating at the ends of $A$ (for a $d=2$-dimensional bulk network, 
or more generally at the boundary of $A$ in a  general network of $d$ bulk dimensions). Thus, $\mathcal{Z}_A^{(nm)}$ represents the partition function of the spin model with twisted boundary conditions leading to an extra spin domain, and $\mathcal{Z}_0^{(nm)}$ represents the partition function with uniform $g=1$ at the boundary.

The difference: $\Delta F^{(nm)} = -\(\log \mathcal{Z}_A^{(nm)} - \log \mathcal{Z}_0^{(nm)}\)$, then represents the free-energy cost of the extra boundary domain (region $A$).
We further note that, in the replica limit, $m\rightarrow 0$, $\Delta F^{(nm)}$ reduces to Eq.~\ref{eq:replica} (the extra logarithm in the free energy is automatically produced by the
replica limit $m\to 0$). Hence, we may interpret the quantum entanglement entropy of the PEPS wave-function as the free-energy cost of a boundary domain in the classical spin model. 

Before analyzing the model in detail, let us briefly comment on the expected behavior in various limits.

\begin{figure}[tb]
\begin{center}
\includegraphics[width=\linewidth]{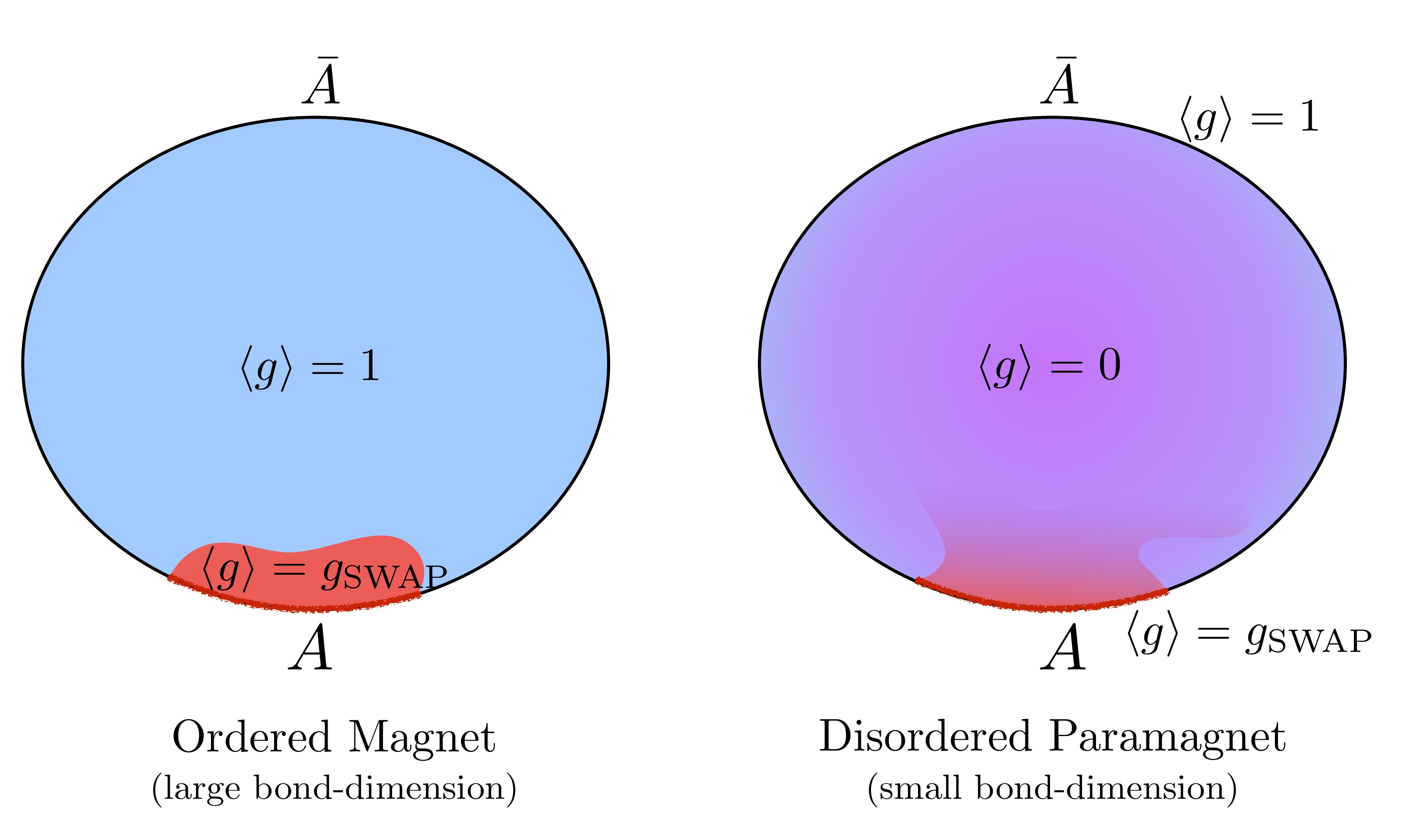}
\vspace{-.1in}
\end{center}
\caption{
{\bf Free energy cost of a boundary twist -- } In the spin-model, the entanglement of a sub-region $A$ on the physical boundary is given by the free energy cost of twisting the boundary conditions from $g=1$ in $\bar{A}$ to $g=g_\text{SWAP}$ inside $A$. In the ordered phase (left-panel), introducing a domain costs extensive energy corresponding to volume law entanglement. In the disordered phase (right-panel), the bulk domain walls are condensed, and there is a free-energy cost only at the edge of $A$, corresponding to area-law entanglement.}
\label{fig:dws}
\end{figure}

At large bond-dimension, we expect an ordered (magnetic) phase of the spin model. Here, there are two cases. First, if the boundary field is weaker than the bulk spin order (i.e. if the logarithm of the bulk bond dimension times the number of nearest neighbors for each boundary spin exceeds the logarithm of the physical on-site Hilbert space dimension), the system will ignore the boundary field and order to $g=1$ everywhere. Ignoring the boundary field inside $A$ will cost an extensive amount of energy, $\sim \log D_p$ per site. This leads  to entanglement scaling like the logarithm of the physical Hilbert space of $A$, corresponding to maximally thermal (i.e.: volume-law) entanglement. If, on the other hand, the spins are ordered, but the bulk spin-exchange is weaker than the boundary fields, then the boundary spins will follow the field, and the minimal free energy cost occurs when the interface between the boundary domain surrounding $A$ follows the geodesic (in $2d$ RTNs, or more generally a minimal-area spanning surface in arbitrary $d$-RTNs) from the left edge of $A$ to the right edge (see Fig.~\ref{fig:dws}). In a regular $2d$ RTN, this geodesic scales as the length $A$, leading to a volume-law behavior of the entanglement entropy. 

In contrast, for low-bond dimension, we expect a disordered (paramagnetic) phase of the spin model, in which the boundary spins align with the boundary field. This paramagnetic phase can be viewed as a bulk ``condensate" of domain walls between fluctuating spin domains. This domain wall condensate can freely absorb the extra domain wall extending from the boundary (see Fig.~\ref{fig:dws}), such that, for a distance exceeding the correlation length from the boundary, there is no additional free energy cost from the boundary domain. In this regime, there will only be a finite energy cost originating from a region of size the bulk correlation length near the spin-twists at the boundary of $A$. In this limit, the free-energy cost of the boundary domain will scale like the boundary of $A$, corresponding to area-law scaling of entanglement. 

In the following, we confirm these expectations by examining the limiting cases of large and small bond dimension within high- and low- temperature expansions of the spin model.

\section{Limiting Cases}

\label{secLimitingCases}
We next compute the entanglement entropy of the random tensor network for two simple limiting cases: low bond dimension (corresponding to the high temperature disordered phase of the spin model) and high bond dimension (corresponding to the low temperature ordered phase of the spin model).

\subsection{High temperature limit (area  law phase)}
For simplicity, we will 
make the approximation that the boundary spins are fixed by the boundary fields to either $1$ or $g_\text{SWAP}$ outside or inside of $A$ respectively. Then, we can perform a high-temperature expansion of the partition function with and without the boundary fixing terms in $A$.

\subsubsection{Leading contribution to area 
law coefficient}
The leading order terms in the high-temperature (low bond dimension) expansion are:
\begin{align}
\mathcal{Z}^{(nm)} &\approx \frac{1}{(Q!)^\Omega}\sum_{\{g\}}\(1+J\sum_{\<ij\>}C(g_i^{-1}g_j)+\dots\)
\nonumber\\
&\approx 1+\frac{J\Omega z}{2} \frac{1}{Q!}\sum_{g}C(g)+\mathcal{O}(J^2),
\end{align}
with  $z$ the coordination number of the graph $G$ (assumed to be constant for simplicity),
and $\Omega = \left| G \right|$ is the network volume (number of sites).


 The 
sum
 $\sum_{g}C(g)$ can be evaluated analytically for arbitrary $Q=mn$ using the function
\begin{equation}\label{eq:F(Q,x)}
F(Q,x)=\sum_{g\in S_{Q}}x^{C(g)}=\frac{\Gamma(Q+x)}{\Gamma(x)}.
\end{equation}
This result can be obtained by solving the recursive equation $F(Q,x)=\sum_{g\in S_{Q-1}}(x^{C(g)+1}+\sum_{k=1}^{Q-1}x^{C(g t_{k})})=(x+Q-1)F(Q-1,x)$ starting from $F(1,x)=x$, where $t_{k}$ denotes the transposition of the elements $k$ and $Q$, which will not change the number of cycles when  multiplied with
 a permutation of the first $(Q-1)$ elements, i.e.\,$\forall g\in S_{Q-1}: C(g t_k)=C(g)$. Then from \eqref{eq:F(Q,x)} we can evaluate $\frac{1}{Q!}\sum_{g}C(g)=\frac{1}{Q!}\partial_xF(Q,x)|_{x\to1}=h_{Q}$ where $h_Q$ is the $Q^\text{th}$ harmonic 
number.\footnote{Recall that
$h_Q = \sum_{k=1}^Q 1/k = \gamma
+
\Gamma'(Q+1)/\Gamma(Q+1)$, where $\gamma= - \Gamma'(1)$, and $\Gamma(x)$ is the usual Gamma Function.}
 In the replica limit
$Q\rightarrow 0$, this becomes $\approx \frac{\pi^2}{6}Q$.

Assembling these results, the leading order high temperature expansion for the bulk partition function is:
\begin{align}
\mathcal{Z}^{(nm)}\approx \(1+\frac{\pi^2}{6}JQ\)^\Omega.
\end{align}
We notice that to lowest order in the high-temperature expansion, the above answer does not depend on boundary conditions, except for the bonds at the boundaries  of the entanglement region crossing from $A$ to $\bar{A}$. These give weight $\delta \mathcal{Z}_A\approx 2J \(C(g_\text{SWAP})-C(1)\) = 2J(m-nm)$. In particular, all other lowest order contributions to the partition function will cancel between $\mathcal{Z}^{(nm)}_A$ and $\mathcal{Z}_0^{(nm)}$ in Eq.~\ref{eq:replica}. The Renyi entropy from this contribution following Eq.~\ref{eq:replica} gives:
\begin{align}
S^{(n)}_A\approx 2J\approx 2\log D_e,
\end{align}
i.e. we obtain
area
law scaling with a contribution of roughly the log of the bond dimension for each bond crossing the boundary.

\subsubsection{Higher order contributions} Corrections to the above expression are generated only at higher orders in the high temperature expansion where one starts to obtain linked clusters that form an arc from a site within $A$ to a site outside of $A$. 

As a concrete example, for a square lattice tensor network, the lowest order contribution comes from a three link cluster starting at the boundary spin within $A$ going up into the bulk over across the boundary of $A$ and back down to $\bar{A}$. Let us label the two bulk sites in this three link loop as $1,2$. Furthermore let us expand the ``number of cycles'' class function onto irreducible characters, $\chi_\xi$ associated with the irreducible representation $V_\xi$ labelled by $\xi$: $C = \sum_\xi \alpha_\xi \chi_\xi$ where $\alpha_\xi$ are coefficients that can in principle be determined for any $n,m$. Then, the leading contribution in the high temperature expansion is:
\begin{align}
\Delta \mathcal{Z}_A&\approx -\frac{1}{(Q!)^2}\frac{J^3}{3!}\sum_{g_1,g_2}C(g_1)C(g_1^{-1}g_2)C(g_2^{-1}g_\text{SWAP})
\nonumber\\
& = -\frac{J^3}{3!}\sum_\xi \frac{\alpha_{\xi}^3}{\(\text{dim} V_\xi\)^2}\[\chi_\xi(g_\text{SWAP})-\chi_\xi(1)\],
\end{align}
up to terms of order $\mathcal{O}(J^4)$.
Although these character sums are difficult to evaluate for general $Q$, as the detailed group structure changes as a function of $Q$, we can readily extract some general features without explicit computation.
First, each term will be positive, since the twist of $g$ from $1$ to $g_\text{SWAP}$ from one end of the arc to the other in $\mathcal{Z}_A$ reduces the amplitude compared to the untwisted one in $\mathcal{Z}$. Next, we can take the crude bound: $C(g)<Q!$ to bound the amplitude of this term by $|\Delta \mathcal{Z}_A|\leq\frac{1}{3!}J^3(Q!)^3$. Hence, to this order, the contribution to entanglement is bounded above by:
\begin{align}
\delta S_A \leq \lim_{Q\rightarrow 0}\partial_Q \frac{1}{3!}J^3(Q!)^3 = \frac{1}{2}J^3+\mathcal{O}(J^4)
\end{align}
For $(JQ!)\ll 1$, the contribution to the high temperature expansion of larger linked clusters will be suppressed exponentially in cluster size in the high temperature limit, showing that only small clusters near the boundary of $A$ contribute appreciably, resulting in area law entanglement. This general structure continues to higher orders, demonstrating the area-law entanglement within the finite radius of convergence of the high-temperature expansion.

\subsection{Low temperature limit (volume law phase)}
From now on we use the simplified notation 
$\mathcal{Z}_A$ for $\mathcal{Z}^{(nm)}_A$,
and $\mathcal{Z}_0$ for $\mathcal{Z}^{(nm)}_0$.

Next, let us consider the opposite limit of large bond dimension (low temperature in the spin model).
 We will primarily consider a regular $2d$ network (e.g. a $2d$ square lattice) and then comment on other geometries. Here, the spin model is deep in the ordered phase, and we can approximate $\mathcal{Z}_0$ and $\mathcal{Z}_A$ (assuming that the bulk bond dimension is much larger than the physical bond dimension, so that the boundary fields  $h_v$  are ineffective at pinning the spins inside $A$ to $g_\text{SWAP}$) by a single configuration $\forall i: g_i = 1$, which has weight $e^{z \Omega J Q/2}$. 


For example, consider a $2d$ square lattice network. The difference between $\mathcal{Z}_A$ and $\mathcal{Z}_0$ for the dominant low-temperature configuration is: $\mathcal{Z}_A-\mathcal{Z}_0 = \(e^{h_AL_A\(C(g_\text{SWAP})-C(1)\)}-1\)\mathcal{Z}_0 = \(e^{h_AL_Am(1-n)}-1\)\mathcal{Z}_0$. Differentiating with respect to $m$ and setting $m$ to zero (recall $\mathcal{Z}_0\rightarrow 1$ in this limit) gives:
\begin{align}
S_A^{(n)}\approx \sum_{v\in A} h_v = L_A\log D_p,
\end{align}
where $D_p$ is the on-site Hilbert space dimension of the physical degrees of freedom (see the line below Eq. \ref{LABELEqBoundaryMagneticFields}).

Thus, the low-temperature (large bond dimension) expansion, gives high-temperature thermal entanglement behavior, in stark contrast to the area-law behavior observed in the high-temperature expansion for low bond dimension. The finite radius of convergence for the high-and low- temperature expansions in the spin-model indicate that there must be a critical point in which the replica spins order, i.e. in which the tensor network wave-function changes from area- to volume- law. Before investigating the properties of this transition for regular $2d$ networks, we briefly comment on other possible tensor network geometries.

\subsection{Other tensor-network geometries}
Besides the regular $2d$ tensor network considered above, the low-temperature expansion can be easily be extended to arbitrary graphs. In each case the low-temperature regime of the effective spin model will be dominated by the spin configuration which has $g_i=g_\text{SWAP}$ for a region including the boundary region $A$ and extending into the bulk into some region whose boundary cuts the minimal number of bonds, and $g_i=1$ everywhere else. The resulting entanglement will then scale as the surface area of the minimal bulk region (see Fig.~\ref{fig:intro}), reproducing the Ryu-Takayanagi formula~\cite{Ryu:2006fj} and agreeing with the infinite $D_e$ results of Hayden et al~\cite{Hayden2016}. For example, regular $d$-dimensional RTNs will have volume-law scaling in all dimensions at sufficiently large $D_e$. Another natural network geometry to consider is that of a ``multiscale renormalization ansatz" or MERA, which describe critical (e.g. conformal field theory, CFT) wave-functions of the boundary degrees of freedom, and provide a discrete regularization of hyperbolic spacetime associated with holographic gravity duals to the boundary CFT. In these networks, the shortest distance curve connecting the end points of a length-$L_A$ boundary region, will have length scaling with $\log L_A$. Hence at large $D_e$ the dominant configuration of the spin model will have a domain between $g = g_\text{SWAP}$ and $g=1$ configurations, whose boundary has length $\sim \log L_A$, costing free energy $\sim \log L_A$, corresponding to the characteristic log-scaling of entanglement for the boundary degrees of freedom.

\section{Universal properties of entanglement phase transitions}
\label{secUniversalFeatures}

By tuning the bond dimension $D_e$, or more precisely by tuning the edge mutual information $0\leq I_e\leq 2 \log  D_e$, the spin model corresponding to $\overline{(\text{tr}\rho^n)^m}$ undergoes a phase transition at a critical point $I_{e,c}=2\log D_c$ (expressed in terms of a critical bond dimension $D_c$ which will generally be non-integer) for each $m,n$. Correspondingly, the entanglement entropy of the random tensor network states switches from area-law to volume law. What are the critical properties of this area-to-volume law transition? To answer this question, we need to confront the task of analytically continuing the parameter $m$ from discrete integer values to a continuous parameter that can be taken to zero. 

\subsection{General features of the critical point}

For simplicity, we will focus on the properties of  a random PEPS defined on a 2d square lattice $G$ --- we will consider the case of random lattices below in Section~\ref{SecRandomGravity}. To summarize the results of the sections above, we can compute the disorder-averaged Renyi entropies $S_n = \frac{1}{1-n} \overline{\log \frac{\tr \rho_A^n}{(\tr \rho)^n}}$ for random tensor networks with bond dimension $D_e$ using the usual replica trick as a difference of free energies
\begin{align}
S_A^{(n)} &= \lim_{m \to 0} \frac{1}{m(n-1)} \left(F_A - F_0 \right).
\end{align}
Here, we have introduced the free energies $F=- \log \mathcal{Z}$, where as before 
$\mathcal{Z}
\to \mathcal{Z}_0= \overline{(\tr \rho^n)^m}$ and $\mathcal{Z}\to\mathcal{Z}_A=\overline{(\tr \rho_A^n)^m}$, and we have used the fact that $\mathcal{Z}_0=\mathcal{Z}_A=1$ in the replica limit $m\rightarrow 0$ to introduce the extra logarithms required to convert powers of
partition functions $\mathcal{Z}_A$\
and $\mathcal{Z}_0$
 to free energies. As we have argued above, these traces can be interpreted as actual partition functions of a statistical mechanics model defined on the graph $G$ with spins $g_v \in S_{Q=nm}$.  The partition function $\mathcal{Z}_0 $ has a boundary field that favors the trivial permutation, while $\mathcal{Z}_A$ has a boundary field favoring $g_{\rm SWAP}$ on the entanglement interval $A$,
and $g=1$ on its complement $\bar{A}$. In the high temperature (low bond dimension) limit, the free  energy cost $F_A - F_0 $ of this domain wall vanishes so that the Renyi entropies satisfy an area law $S_n \sim {\rm cst}$, while a 
low temperature expansion simply gives $\mathcal{Z}_A \sim \mathcal{Z}_0 D_p^{m(1-n) L_A}$  to leading order, where $L_A$ is the
length  of the interval $A$, implying a volume law scaling $S_n = (\log D_p) L_A$ deep in the ordered ($D_e \to \infty$) phase. (Recall that $D_p$ is the dimension of the physical Hilbert space.) We 
expect\footnote{at least for sufficiently small values of $Q=nm$}
 a critical point separating these two phases for a critical coupling $J_c(nm)$ (with $J=\log D_e$), and our goal is to keep track of the universal properties of this transition as $m \to 0$.

A simple point of our model is given by $Q=nm=2$ which corresponds to the two-dimension Ising model. In general, assuming that the transition is of second order, it should be described a by Conformal Field Theory (CFT). Since in the replica limit $m \to 0$, the partition functions become unity $\mathcal{Z}_0 = \mathcal{Z}_A=1$, we know that the central charge of the CFT in that limit is $c=0$. (This is because $c$ measures the way the free energy changes when a finite scale is introduced: since here the partition function is trivial
for any finite size system, this immediately implies that $c=0$.) Since the only unitary CFT with $c=0$ is the trivial theory with a single
identity operator, we know that the CFT we are after is in a class of non-unitary Logarithmic CFTs (LCFT)~\cite{GURARIE1993535,0305-4470-35-27-101,doi:10.1142/9789812775344_0032,1751-8121-46-49-494001}, where the non-unitarity leads to the appearance of the logarithmic correlations even at the critical point -- we will come back to this point below. Importantly, since the bulk properties of the transition only depend on the product $Q=nm \to 0$, the location of the bulk transition point as well as all
$Q \to 0$ bulk critical exponents are the same for all Renyi entropies
in the limit $Q \to 0$. (This includes in particular the
correlation length exponent $\nu_{Q} \to \nu$  which is thus the same for all Renyi entropies in the limit $Q \to 0$). This is in sharp contrast with the construction of Hayden et al. deep in the ordered phase\cite{Hayden2016}, where different Renyi entropies correspond to distinct classical spin models, 
and are expected to have transitions (possibly of  first order) at different values of  $D_e$.

To analyze the scaling of the entanglement entropy at the critical point, we note that the ratio of partition functions  $\mathcal{Z}_A/\mathcal{Z}_0$ that appears in the free energy difference $F_A - F_0 = - \log \frac{\mathcal{Z}_A}{\mathcal{Z}_0}$, corresponds in the CFT language 
to the two-point function of boundary condition changing (BCC) operators~\cite{cardy_boundary_2006,cardy_conformal_1984}. Note that strictly speaking, the boundary fields break conformal invariance but on large distances we expect these fields to flow to infinity, corresponding to conformally invariant fixed boundary conditions where the spins are pinned to the identity perturbation or to $g_{\rm SWAP}$. Introducing the operator $\phi_{\rm BCC}$ that implements  this change of boundary condition, $\mathcal{Z}_A/\mathcal{Z}_0 = \langle \phi_{\rm BCC}(L_A) \phi_{\rm BCC}(0) \rangle$ where the operators are inserted at  the boundary of the entanglement interval $A$.  In general, there are non-universal extensive terms corresponding to the difference of boundary free energies associated with the different boundary conditions, but this difference vanishes by symmetry in our case since our two boundary conditions where the spins are fixed to different values are related by symmetry. As an instructive example, let us briefly discuss the Ising model case $Q=nm=2$ (say, $m=1$ and $n=2$): then $\mathcal{Z}_A/\mathcal{Z}_0$ can be mapped by duality to the spin-spin correlation function in the dual model.  In the ordered phase, the spin-spin correlation function in the dual model is disordered and decays exponentially, leading to a volume-law contribution to the entanglement, whereas in the disordered phase, the dual model has long-range order so the spin-spin correlation function 
tends to a  constant at large separation, leading to area law entanglement. This can of course be interpreted in terms of domain wall free energy cost. The corresponding BCC operator that implements a change between boundary conditions where the spins are pinned down to opposite directions has scaling dimension $\Delta = \frac{1}{2}$. Going back to the replica limit, we find that the Renyi entropies scale as
\begin{equation}
S_A^{(n)} =   \frac{2}{n-1}
\lim_{m\to 0} {1\over m} \Delta(n,m)
\ \log L_A + {\rm cst},
\label{eqBCCRenyi}
\end{equation}
where $\Delta(n,m)$ is the dimension of the BCC  operator
associated with the boundary field in the entanglement interval $A$ in the CFT describing the critical point. Note that because the change in boundary conditions becomes trivial for $m=0$ and/or $n=1$, we expect $\Delta(n,m)$ to vanish in these limits, making~\eqref{eqBCCRenyi} well defined.

 Away from the critical point, the two point function of the BCC operators should scale as $C_{n,m}/L_A^{2 \Delta (n,m)} f_{n,m} (L_A/\xi_{nm})$, leading to the following scaling form for the Renyi entropies by using the replica trick
\begin{equation}
S_A^{(n)}  =   \frac{2}{n-1} \left.\frac{\partial \Delta}{\partial m} \right|_{m=0} \log L_A + C_n + f_n \left( \frac{L_A}{\xi}\right),
\label{eqEntanglementLog}
\end{equation}
with $\xi \sim \left| D - D_c\right|^{-\nu}$ the correlation length in the limit $Q=nm \to 0$, $f_n$ some universal scaling functions and $C_n$ some non-universal constants. In order to isolate the universal contributions, one can also take the derivative with respect to $\log L_A$:
\begin{equation}
L_A  \frac{\partial S_A^{(n)} }{\partial L_A} =   g_n \left( \frac{L_A}{\xi}\right),
\end{equation}
with $g_n(0) = \frac{2}{n-1} \left.\frac{\partial \Delta}{\partial m} \right|_{m=0}$.

\subsection{Analytical continuation of the spin model}

\begin{figure}[tb]
\begin{center}
\includegraphics[width=0.75\linewidth]{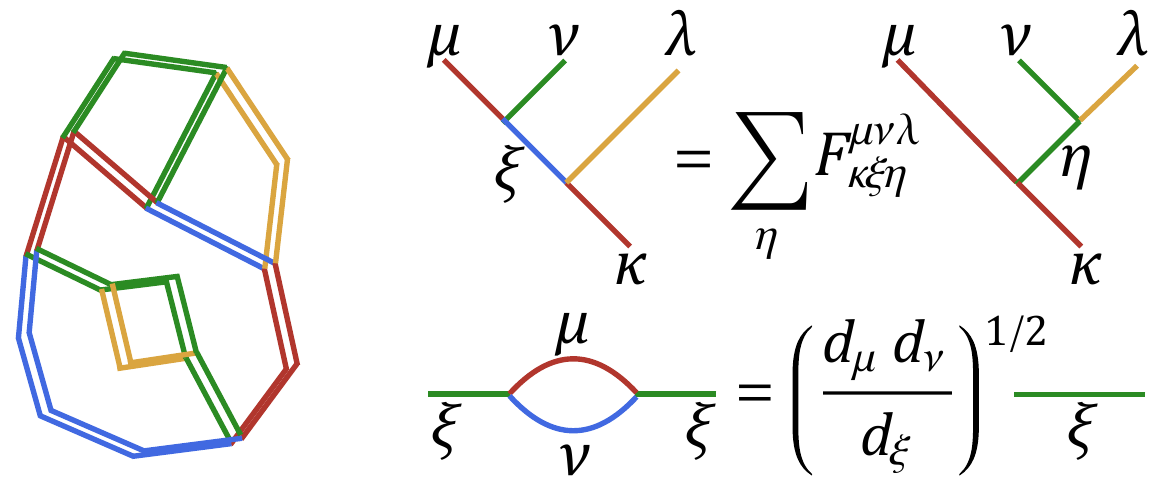}
\vspace{-.1in}
\end{center}
\caption{
{\bf Loop model --} Admissible diagrams (string-net configurations) are double strings satisfying the fusion rules. In each layer, the string-net amplitude $\Phi$ is independently evaluated following the local moves listed on the right.}
\label{fig:stringnet}
\end{figure}

To identify the relevant CFT in the replica limit $m \to 0$, we first need to analytically continue our spin model to real values of $Q=n m$. To that end, we first map the classical spin model onto a loop model (high temperature expansion). First note that the function $C(g)$ that counts the number of cycles in the permutation $g \in S_Q$ is a class function as it satisfies $C(g) = C(h^{-1} g h)$ for all $h \in S_Q$. A complete basis for such class functions is given by the characters of the group $ \chi_\xi (g) = {\rm tr}_{V_\xi} g$ given by the trace of the representation of the group element $g$ in the irreducible representation (irrep) $V_\xi$, where we used $\xi$ to label irreps. For reasons that will become clear below, we define $\overline{\chi}_\xi (g) = \chi_\xi (g) \frac{{\rm dim} V_\xi}{Q!} $. By decomposing the Boltzmann weights onto these (modified) irreducible characters, we write
\begin{equation}
\mathcal{Z}_0 \propto \sum_{\lbrace g_v \in S_{Q}\rbrace} \prod_{\langle v,v^\prime \rangle} \left(1 + \sum_{\xi \neq 1} K_\xi \overline{\chi}_\xi (g_v^{-1} g_{v^\prime}) \right),
\label{eqLoopModel}
\end{equation}
where $\xi \neq 1$ labels irreps different from the trivial one.
 The couplings $K_\xi $ can be computed 
exactly for our microscopic interaction ${\rm e}^{J C(g_v^{-1} g_{v^\prime})}$, 
\begin{equation}
K_\xi=\frac{H_\xi}{H_1}, H_\xi=\prod_{(i,j)\in Y_\xi}({\rm e}^J+j-i),
\end{equation}
where $Y_\xi$ denotes the Young diagram of the irrep $\xi$ and $(i,j)$ is the box at row-$i$ and column-$j$ in $Y_\xi$ (with the indices $i,j=0,1,2\cdots$ start from zero).
Nevertheless the precise expression is not important in what follows. Since we are interested in universal properties, we will consider a more general model where the couplings $K_\xi$ are arbitrary. 
By expanding this product and summing over the spins, we get a high temperature expansion with loops carrying a label (or color) associated with the irrep $\xi$, as illustrated in Fig.\,\ref{fig:stringnet}. Diagrams with ``tadpoles'' are forbidden since $\sum_g \chi_\xi(g) = 0$, and different irreps cannot propagate in the same loop thanks to the convolution formula
\begin{equation}
\sum_{h\in S_Q} \overline{\chi}_\xi(g h^{-1}) \overline{\chi}_{\xi^\prime}(h) = \delta_{\xi, \xi^\prime} \overline{\chi}_\xi(g).
\end{equation}
These loops can branch and cross, with a condition related to the decomposition of the tensor products of the associated irreps containing the trivial representation.

To specify the general weights of diagrams, we evoke the mathematical structure of the representation category  of the
symmetric group, $\text{Rep}_{S_Q}$,
which is the fusion category formed by the irreps of $S_Q$. 
All admissible diagrams $X$ that contribute to the partition function $\mathcal{Z}_0$ satisfy the fusion
(= Clebsch-Gordon)  rules of $\text{Rep}_{S_Q}$. The loop model is given by
\begin{equation}
\mathcal{Z}_0=\sum_{X}\bigg(\prod_{\xi}K_{\xi}^{|X_{\xi}|}\bigg)\Phi(X)^2,
\label{eq:loopstringnet}
\end{equation} 
where $|X_{\xi}|$ is the total length of the $\xi$-type string in the diagram $X$ and $\Phi(X)$ is the string-net amplitude of $X$. The couplings $K_{\xi}$ set the string tension and the remaining part $\Phi(X)^2$ is topological. The amplitude $\Phi(X)$ is uniquely specified by local constraint equations listed in Fig.\,\ref{fig:stringnet}, where $F^{\mu\nu\lambda}_{\kappa\xi\eta}$ denotes the $F$-symbol 
(6j-symbol) of $\text{Rep}_{S_Q}$ and $d_\xi\equiv{\rm dim}V_\xi$ is the dimension of the irrep $\xi$. Given a string-net configuration $X$, one can always use the local transformations to deform $X$ to the trivial configuration with no strings. Then the coefficients accumulated along the path of deformation give the amplitude $\Phi(X)$. For the symmetric group  $S_Q$,
all $F$-symbols of the irreps can be chosen real, so the amplitude $\Phi(X)$ is also real and its square $\Phi(X)^2$ is positive. By introducing an unnormalized string-net wave function $|\{K_{\xi}\}\rangle=\sum_{X}\prod_{\xi}K_{\xi}^{|X_{\xi}|/2}\Phi(X)|X\rangle$, with
$|X\rangle$ an orthonormal basis,
 the partition function $\mathcal{Z}_0$ can be expressed as the
squared norm of the unnormalized string-net wave function, i.e.\,$\mathcal{Z}_0=\langle\{K_{\xi}\}|\{K_{\xi}\}\rangle$. Driven by the set of couplings $\{K_\xi\}$, the string-net state $|\{K_{\xi}\}\rangle$ can undergo a deconfinement-confinement transition, which corresponds to the disorder-order transition in the $S_Q$ spin model.


\subsection{Truncation of the loop model: percolation universality class}

The above high temperature expansion leads to a complicated model of branching loops (``nets''), with one flavor of loop per non-trivial irreducible representation. This loop model is very complicated in general, and is still hard to analytically continue to $m=0$ since the number of loop flavors depends on $Q$. To make further progress, we will therefore consider simpler versions of that model by modifying the couplings $K_\xi$. We will consider an especially simple choice of couplings that makes the model tractable, but that also accidentally enlarges the symmetry of the model. We will then study the stability of 
the fixed point we find in the model with simplified couplings, to symmetry-breaking perturbations that should appear for the generic
 physical couplings corresponding to the spin model~\eqref{eqSpinModel}, and infer from there the fate of the true infrared (IR) fixed point.

 Our first step is to consider a specific choice for the couplings $K_\xi$. A simple choice is provided by considering uniform couplings over all characters $K_\xi=K$, in which case eq.~\eqref{eqLoopModel} can be simplified using the identity $\sum_\xi \overline{\chi}_\xi(g )= \delta_{g,1} $. In that case, the partition function reduces to
\begin{equation}
\mathcal{Z}_0 = \sum_{\lbrace g_v \in S_{Q}\rbrace} \prod_{\langle v,v^\prime \rangle} \left(1 -K + K \delta_{g_v, g_{v^\prime}}\right),
\label{eqPotts}
\end{equation}
up to normalization factors that tend to one in the $Q \to 0$ limit. This coincides with the high temperature expansion of a Potts model with $Q! = \left| S_{Q} \right|$ states. The product in this partition function can be expanded onto the so-called Fortuin-Kasteleyn clusters~\cite{FORTUIN1972536} where an edge is occupied with weight $K$, not occupied with weight $1-K$, and where each connected cluster (including single sites) carries a weight $Q!$ (the number of states). In the replica limit $Q \to 0$, this maps onto a bond percolation problem where $K$ is the probability for a bond to be occupied. The partition function in that limit is $\mathcal{Z}_0=1$ as needed, and the CFT corresponding to the critical point $K_c=\frac{1}{2}$ on the square lattice has central charge $c=0$ as required by the general considerations mentioned above. The correlation length diverges as $\xi \sim \left| K - K_c\right|^{-4/3}$ at the transition~\cite{PhysRevB.23.429}.

\subsection{Stability of the percolation fixed point}

Whereas the above percolation fixed point could be a reasonable candidate for the transition in the replica limit, it is important to notice that our choice of couplings $K_\xi=K$ enlarged the symmetry of the model dramatically. Starting from a model with $S_Q$ symmetry (or $S_Q \times S_Q$ corresponding to right and left multiplication), we obtained a model with a much larger $S_{Q!}$ symmetry. To study if this enlarged symmetry is emerging at the critical point for generic couplings, we study the stability of this $S_{Q!}$ fixed point against including $S_Q$-symmetric perturbations that break the $S_{Q!}$ symmetry. 

We first recall the formulation of the Potts field theory with $Q!$ states. Let $\phi_a$ denote the coarse-grained order parameter of the Potts model obtained as the scaling limit of the magnetization operator. It has $\left ( Q!-1\right )$ components as it satisfies $\sum_{a=1}^{Q!} \phi_a = 0$. In terms of this order parameter, one can write an effective Landau theory 
\begin{equation}
\label{eqLandauPotts}
{\cal L}_{\rm Potts} =\frac{1}{2} \sum_a (\partial_\mu \phi_a)^2 + \frac{m^2}{2} \sum_a \phi_a^2 + g \sum_a \phi_a^3 + \dots ,
\end{equation}
with the crucial constraint $\sum_{a} \phi_a = 0$. For $Q! < 2$, a perturbative RG analysis near the upper critical dimension $d_{\rm uc} = 6$ shows the existence of a non-trivial fixed point describing a second-order phase transition. Note that for $Q!=2$, $\sum_a \phi_a^n$ vanishes for any odd $n$ (and $n=3$ in particular) because of the $S_2=\mathbb{Z}_2$ symmetry of the Ising model. In two dimensions, the operator content at the critical point is well-known~\cite{diFrancesco1987}: (1) There are thermal perturbations that transform trivially under $S_{Q!}$, among which the first thermal operator is relevant and drives the transition (it has scaling dimension $\Delta_\epsilon = \frac{5}{4}$ for $Q=0$); (2) There are magnetic operators that transform as $[Q!-1,1]$ under $S_{Q!}$ where  $[Q!-1,1]$ denotes the Young tableau with 2 rows having
$(Q!-1)$ boxes in the first row and one box in the second row; (3) Finally there are the so-called $M$-hull operators with $M \geq 2$ integer, which can be thought of as creating $M$ Fortuin-Kasteleyn clusters (percolation clusters for $Q=0$) -- they transform under the symmetry as $[Q!-M,M]$.

This $Q!$-state Potts model has a $S_{Q!}$ symmetry where  the action of $S_{Q!}$ onto the fields $\phi_a$ is given by $\phi_a \longrightarrow \phi_{g(a)}$, where $a=1,\dots,Q!$ are some Potts colors (or states) and $g \in S_{Q!}$ is a permutation of these colors. We want to include perturbations that break this $S_{Q!}$ symmetry down to the subgroup $S_Q \subset S_{Q!}$ defined by the action $\phi_a \longrightarrow \phi_{h a}$, where we now think of $a \in S_Q$ as being an element of $S_Q$ and $h \in S_Q$ acts on $a$ by left multiplication. It is straightforward to show that the leading perturbation implementing the symmetry breaking $S_{Q!} \longrightarrow S_Q$ is given by
\begin{equation}
{\cal L} ={\cal L}_{\rm Potts} + \sum_{a,b \in S_{Q}} W(a^{-1} b) \phi_a \phi_b + \dots
\end{equation}
where $W$ is a class function of $S_Q$. Again, crucially the labels $a,b$ are now interpreted as elements of the group $S_Q$. The only allowed function $W(a^{-1} b)$ that would respect the $S_{Q!}$ symmetry is $\delta_{a,b}$, but any class function of $S_Q$ is enough to satisfy the $S_Q $(in fact  more precisely, the  $S_Q \times S_Q$) symmetry. The perturbations $\phi_a \phi_b$ with $a\neq b$ are therefore allowed by $S_Q \times S_Q$ even if they break $S_{Q!}$, and they should generally be included in the action. They correspond to the two-hull operator mentioned above: they have scaling dimension $\Delta_{\rm 2-hull}=\frac{5}{4} < 2$ at the percolation fixed point (in the limit $Q \to 0$) and they are therefore relevant. We conclude that the percolation fixed point for $Q \to 0$ is generically unstable and thus fine-tuned, and flows to a different fixed point in the infrared (IR).

\subsection{Fate of the perturbed percolation fixed point}

Even if the percolation fixed point with enlarged $S_{Q!}$ symmetry (in the replica limit $Q \to 0$) is unstable, we anticipate that it will provide a useful starting point to analyze the IR fixed point.  The quantum field theory describing the IR fixed point is among a class a notably complicated CFTs with $c=0$  called Logarithmic CFTs~\cite{GURARIE1993535,0305-4470-35-27-101,doi:10.1142/9789812775344_0032,1751-8121-46-49-494001}, which are poorly understood and for which a classification is still lacking. From the above analysis, we know that it corresponds to a Potts model with $Q!$ states perturbed by a two-hull operator, in the limit $Q \to 0$ -- preserving the symmetry group $S_Q \times S_Q$. The fate of the percolation critical point perturbed by a
 two-hull operator has been studied recently~\cite{PhysRevLett.90.090601,2013PhRvB..87r4204N} in a different context, and it was found that
in that case  the critical point  fans 
out into a gapless ``Goldstone'' phase~\cite{PhysRevLett.90.090601}, where there are Goldstone modes living on the sphere $S^{N-1}=O(N)/O(N-1)$ in the limit $N\to 1$~\footnote{Note that the existence of this Goldstone phase in two dimension may seem surprising, but the Mermin-Wagner theorem does not apply in the replica limit $N \to 1$ where the theory is effectively non-unitary, so that symmetry-breaking of $O(N)$ is allowed.}. Although this result was derived in a different context (in particular with a pattern of symmetry breaking that is very different from our case), this
may perhaps suggest
 the intriguing possibility of having an intermediate gapless phase in our model as well. It would be interesting to investigate this limit further, possibly by using a different truncation of eq.~\eqref{eq:loopstringnet} (for example, restricting to a single character). 
We leave a detailed analysis of the field theory of the critical point on regular planar graphs for future work.

\subsection{Randomness in the bond dimension}
We have so far considered the simplifying case of fixed bond dimension $D_e$, which is an integer for each edge. In contrast, the critical bond dimension $D_c$ corresponding to the area-to-volume-law transition will generically be non-integer. There are two possible schemes to access this entanglement transition in random tensor network states. One way is to consider replacing the maximally entangled pair on each edge by a generic entangled pair, as specified in Eq.~\ref{eq: Ie state}. In this way, the ``effective'' bond dimension  will be continuously tunable by varying the bond mutual information $I_e^{(n)}$. A similar setup was also discussed in Ref.~\onlinecite{2018PhRvB..97d5153Y} to allow the application of machine learning techniques to find the optimal RTN representation of given entanglement features. Another way to access  the entanglement transition is to consider inhomogeneous and random bond dimension, whose typical value is equal to $D_c$: $\overline{\log D_e} = \log D_c$. If the fluctuations in $D_e$ are sufficiently weak, we can address their effect on the critical properties of the model perturbatively. Assuming that the bond dimensions are log-normally distributed, i.i.d. on each bond (corresponding to i.i.d. Gaussian randomness of the spin couplings in the equivalent stat-mech model), then by the usual Harris criterion, the bond dimension disorder is an irrelevant perturbation when $\nu >1$ (for a 
tensor network defined on a regular two-dimensional lattice $G$). 

\section{Random geometry and quantum gravity}
\label{SecRandomGravity}

\subsection{Universality and phase transitions for a random geometry}

While considering random PEPS living on fixed regular graphs is interesting, we do not know the actual geometry of the bulk tensor network that would correspond to physical entanglement transitions, including the MBL transition for example. Following the random matrix theory logic, it is natural to take that geometry also  to be random: if universal features remain for completely random networks defined on random planar graphs, there are likely to describe universal features of entanglement transitions in more realistic many-body quantum systems.  More precisely, we will consider a random tensor network whose vertices live on a random trivalent graph -- the trivalent nature of the graph will be 
irrelevant  for the phase transition. We will also restrict to graphs with the topology of a disk so that they have a proper boundary (Euler characteristic $\chi = 1$). Instead of working with a fixed random graph $G$ with $N=\left| G\right|$ vertices, it is convenient to work in the ``grand canonical'' ensemble, and to introduce a chemical potential $\Lambda$ conjugated to the mass of the graph $N=\left| G\right|$ (also called cosmological constant for reasons that will become clear below). Following the replica trick strategy developed above, the entanglement entropy averaged over an ensemble of random graphs can be inferred from the analytic continuation $m \to 0$ of a spin model now defined   on fluctuating
graphs
%
\begin{equation}
\mathcal{Z}_{\rm random} = \sum_{G}  \frac{1}{S(G)} {\rm e}^{- \Lambda \left| G\right|} \mathcal{Z}^{(nm)} (G),
\end{equation}
where $S(G)$ is the symmetry factor of the graph $G$. We are only interested in ``quenched'' quantum gravity where the random graphs are generated with a probability distribution independent of the statistical mechanics model. Note that since $\mathcal{Z}_{S_{nm}}=1$ in the replica limit $m = 0$, there is in fact  no difference between quenched and annealed averages over the random graphs for the cases of interest to us. The critical behavior of the Renyi entropies averaged over disorder of the Random Tensor Network, as well as over 
fluctuating   graphs,  can therefore be obtained directly from the replica limit of our $S_{nm}$ Stat.  Mech.  model defined on annealed fluctuating lattices. In particular, there is no back-action of the ``matter'' (= the Stat. Mech. model) on the random gravity in the replica limit.   The coupling to fluctuating graphs leads to a new critical point which can be interpreted as coupling the CFT obtained on a regular graph (say, a square lattice) to two-dimensional quantum gravity~\cite{KAZAKOV1985295,KAZAKOV1986140,BOULATOV1986641}. Formally, if $S_{\rm CFT}[g_{\mu\nu},\phi,m]$ denotes the action of the CFT in a fixed background metric $g_{\mu\nu}$ perturbed by the mass term $m$ (coupling to the energy operator tuning through  the transition), the partition function on fluctuating lattices can be expressed as~\cite{doi:10.1142/S0217732387001130,KNIZHNIK:1988aa,DAVID:1988aa} 
\begin{equation}
\mathcal{Z}_{\rm random} \approx \int {\cal D} \phi {\cal D} g {\rm e}^{- \int d^2 x \sqrt{g}  \left( \Lambda - \frac{\gamma}{4 \pi} R \right) - S_{\rm CFT}[g_{\mu\nu},\phi,m] },
\end{equation}
where $\Lambda$ is once again our cosmological constant weighting the mass $ \int d^2 x \sqrt{g} $ of the ``universe'', and $\frac{1}{4 \pi}\int  R = \chi $ is a topological term with $\chi$ the Euler characteristic of the graph (Gauss-Bonnet theorem), 
where  $\chi=1$ in our case since we are considering planar graphs with the topology of a disk. The Hilbert-Einstein part of the action is therefore topological in two dimensions. 

Upon 
tuning the cosmological constant $\Lambda$ to obtain an infinite graph (thermodynamic limit), a phase transition can be induced by
simultaneously tuning the coupling $m$ to the transition. This new critical point can be interpreted as a new CFT ``dressed'' by quantum gravity, and the new scaling dimensions are given by the so-called Knizhnik-Polyakov-Zamolodchikov (KPZ) formula~\cite{KNIZHNIK:1988aa}. 
In particular, the new dimension (=conformal weight)  of the energy operator is given by
\begin{equation}
\tilde{h}_{\epsilon} = \frac{\sqrt{1 + 24 h_{\epsilon}}-1}{4},
\label{eqKPZ}
\end{equation}
in the new $c=0$ CFT dressed by quantum gravity, where $\nu = \frac{1}{2-2  h_{\epsilon}}$ is the correlation exponent, and $h_\epsilon$ is the energy operator dimension (=conformal weight)  for the CFT in a flat background. This formula can be derived purely from a field theory framework, and it was checked against the exact solution of various statistical mechanics models defined on random graphs. 

To determine the relevant exponent for our physical system at the boundary, recall that we are interested in the ``canonical ensemble'' where the number of sites $N=\left|G \right|$ is fixed and going to infinity in the thermodynamical limit. From the quantum gravity results, we expect some finite size scaling near the transition in terms of the dimensionless quantity $(D_e-D_c) N^{1/(\tilde{\nu} d_F)} $ where $\tilde{\nu} d_F = \frac{1}{1 - \tilde{h}_{\epsilon}}$, and $d_F$ is the fractal dimension of the random graph $G$. The fractal dimension factor $d_F$ comes
from  the fact that the random graph $G$ is typically very ``spiky'', and the relevant fractal dimension in our case is $d_F=4$ --- this is the fractal dimension of ``pure'' gravity without matter, appropriate since our CFT has central charge $c=0$. Letting $L$ be the linear extent of our system (defined in terms of geodesics), the number of vertices scales as $N \sim L^{d_F}$. Despite this anomalous fractal dimension, it can be shown
 that even
for  these random graphs, one still has $|\partial G| \propto | G|^{1/2}$, where $|\partial G|$ is the number of boundary vertices of $G$. This means that if we let $L=|\partial G|$ be the size (= number of vertices) of our physical spin model living at the boundary of the random tensor network, we have $N \sim L^2$ and the relevant finite size scaling variable in terms of the number of spins at the boundary is $(D_e-D_c) L^{2/(\tilde{\nu} d_F)}$. We therefore identify the physical correlation length defined in units of the number of boundary spins as 
\begin{equation}
\xi_\star \sim \left| D_e - D_c \right|^{-\nu_\star},
\end{equation}
with
\begin{equation}
\nu_\star = \frac{\tilde{\nu} d_F}{2} = \frac{2}{5 - \sqrt{25- \frac{12}{\nu}}},
\label{eqnustar}
\end{equation}
where we have used the KPZ formula~\eqref{eqKPZ} with $\nu$ the correlation length exponent on a regular graph. If the bulk theory is controlled by the percolation fixed point with $\nu=4/3$, this yields $\nu_\star=2$. Interestingly, this saturates the  Harris bound $\nu_\star \geq 2$ for the disordered physical $1d$ boundary spins. We conjecture that the correlation length exponent at the true IR fixed point on a regular lattice satisfies $\nu>\frac{4}{3}$ (which is typical for ``quantum percolation'' problems), which implies that  $\nu_\star \geq 2$ on a random lattice so that the Harris bound is satisfied.

\begin{figure}
\begin{center}
\includegraphics[width=\linewidth]{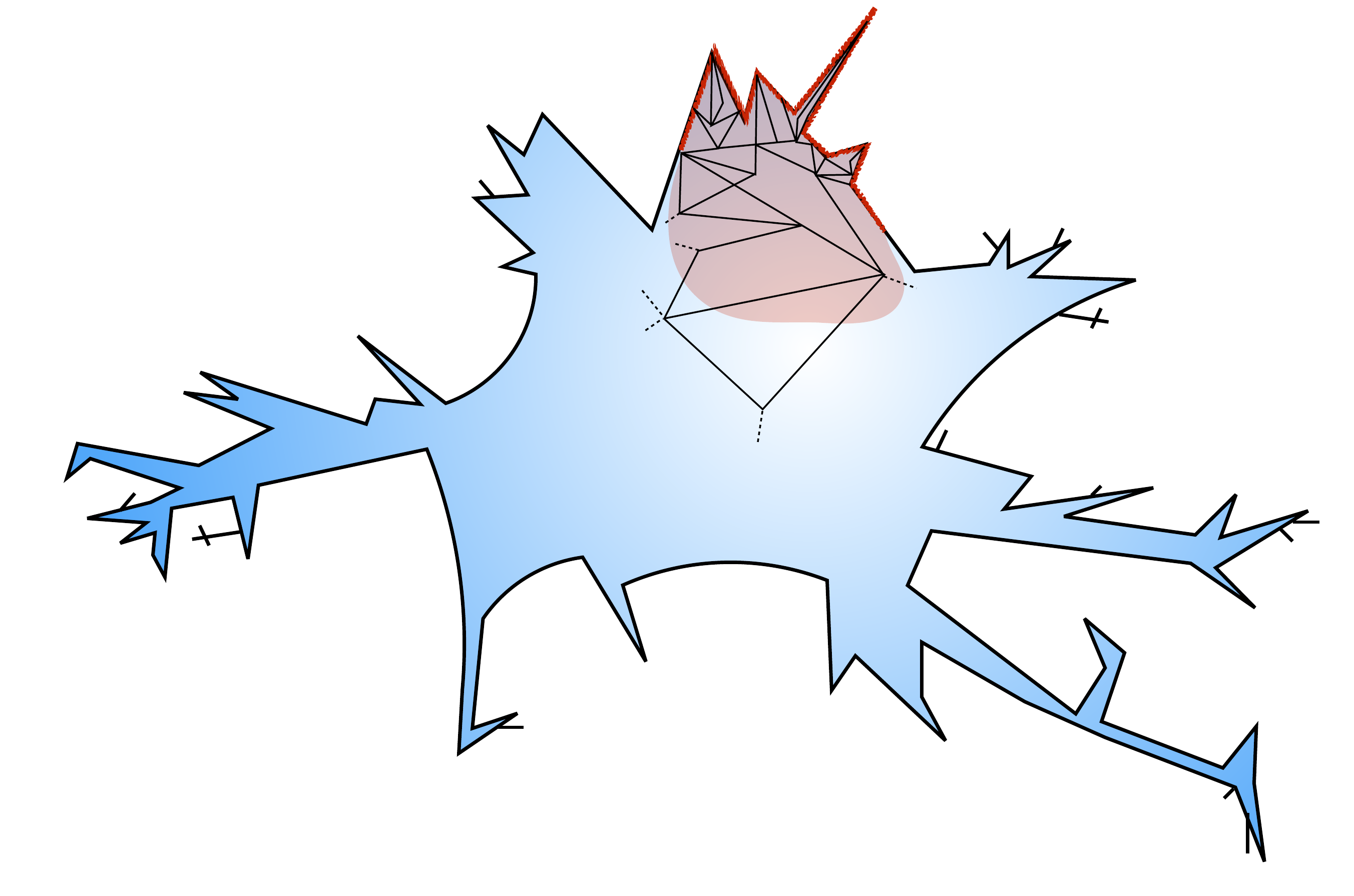}
\vspace{-.5in}
\end{center}
\caption{
{\bf  Random graph with a fractal boundary -- } A schematic example of a random tensor network (blue region), with a fractal boundary. In the ordered phase of the corresponding spin model, the boundary domain (red area) of the entanglement region (thick red line), has a boundary that scales as a sub-extensive power of the number of sites in the entanglement region. This results in unusual power-law scaling of entanglement intermediate between volume- and area-law.
}
\label{fig:randomgravity}
\end{figure}

\subsection{Non-thermal entanglement scaling}

Besides changing the universality class of the transition and renormalizing  the critical exponents, the random geometry also has important consequences for the scaling of entanglement in the large bond dimension (ordered) phase. Recall that at large bond dimension, entanglement is determined by geodesics: if we consider an entanglement interval of size $L_A$ sites at the boundary, in the limit $D_e\to \infty$ the entanglement is given by the Ryu-Takayanagi formula~\cite{Ryu:2006fj}, that is, by a minimal cut through the tensor network --minimizing the energy cost of the domain wall in the spin model language. On a regular (say, square) lattice, this geodesic scales as the size of the interval $\sim L_A$, but on a random lattice this scaling is very different. In particular, recall that  if one considers a subsystem containing $n$  vertices, 
a geodesic joining the boundaries of this box scales only as $n^{1/d_F}$ with $d_F=4$, instead of $\sqrt{n}$ for a regular 2D lattice. Similarly because $N \sim L^2$ with $L$ the number of boundary spins, the boundary of the random graph $G$ has fractal dimension $d_F^b=2$, which implies that for a boundary interval of length $L_A$, the minimal cut of the tensor network (geodesic) scales as $\sqrt{L_A}$ (Fig.~\ref{fig:randomgravity}). This implies that the entanglement has a power-law  scaling in the large bond dimension phase
\begin{equation}
S_A \sim \sqrt{L_A},
\end{equation}
corresponding to a non-ergodic phase since this scaling violates ETH. Random tensor networks defined on random geometry can therefore be used to obtain a power-law scaling of entanglement. In fact, the fractal dimension of the random geometry can be tuned by weighting differently the random graphs (recall that above, we chose a uniform measure consistent with the fact that the Einstein-Hilbert action is trivial in two dimensions). In the field theory language, this can be interpreted as adding background matter fields that have a non-trivial back-action on the random geometry,  thus modifying the fractal dimension. It should also be possible to change the measure over the random graphs -- corresponding to adding ``matter'' at the boundary -- to modify the relation $|\partial G| \propto | G|^{1/2}$ for example. This can be used as a knob to change the nature of the entanglement transition. We leave the study of the entanglement properties of random tensor networks defined on such weighted random graphs for future work.

\section{Discussion}
Our approach uncovers a new class of phase transitions between quantum states with sharply different entanglement scaling. Employing a replica trick enables us to obtain analytic results for arbitrary bond dimension, which in the holographic language corresponds to strong quantum gravity fluctuations in the bulk geometry. Although we defer a detailed analytic understanding of the CFT describing the transition for RTN on regular planar graphs to future work, we emphasize that our approach allows us to formulate a statistical mechanics and field theoretical formulation of such entanglement transitions in any dimension and/or geometry. A promising future application of these ideas could be to examine quantum dynamics under random unitary circuits, away from the large-$N$ limit, which limited previous studies~\cite{PhysRevX.7.031016,2017arXiv170510364N,2017arXiv170508975N,2017arXiv171009835K,2017arXiv170508910V,2017arXiv171009827R,2017arXiv171206836C,2018arXiv180409737Z} to maximally scrambling systems, whose dynamics are effectively classical and mean-field like by construction (see~\cite{2017arXiv171206836C,2018arXiv180409737Z} for recent progress in that direction).{For RTN, our results also provide $1/N$ quantum corrections to the Ryu-Takayanagi formula which would be interesting to compare to the gravity expectations. 

Before concluding, we examine the relationship between the properties of the area-to-volume law entanglement transition in RTN states, to the many-body (de)localization transition between MBL and thermal states. The most notable distinction is that whereas the RTN states exhibit a continuous second order transition between area and volume states, the entanglement has been predicted to jump discontinuously across the MBL transition~\cite{PhysRevX.7.021013,2017arXiv170104827D}. The discontinuity for the direct MBL-to-thermal transition necessarily arises since entanglement of any sub-interval of a thermal system must be extensive, with the volume-law coefficient equal to the thermodynamic entropy density~\cite{2014arXiv1405.1471G}. However, infinitesimally on the MBL side of the transition, the entanglement must be sub-thermal, requiring a discontinuity~\cite{PhysRevX.7.021013}. In contrast, the RTN entanglement entropy near the area-to-volume law transition behaves like a more typical observable, exhibiting a continuous cross-over as the entanglement sub-region length passes through the correlation length of the transition. In fact, the absence of a discontinuity in the volume-law component of the entanglement across the RTN transition signals that the volume-law phase of the RTN states is not fully thermal, and does not satisfy ETH~\cite{2014arXiv1405.1471G}. Exploring the nature of such states which are volume-law entangled, and whether they are related to the eigenstates of a particular class of parent Hamiltonians presents an interesting challenge for future work. 

After the first version of this manuscript appeared on ArXiv, a possibly related entanglement transition in random unitary circuits with projective measurements was proposed in Refs.~\cite{PhysRevX.9.031009,PhysRevB.98.205136,PhysRevB.99.224307,2019arXiv190108092L}. The scaling observed numerically at that transition appears to be compatible with a dynamical exponent $z=1$ and conformal invariance. This suggests that this measurement-induced transition can be described by a (replica) statistical mechanics model similar to what we discussed above. It would be interesting to determine the relation to the entanglement transition in RTN in future works.

It would be interesting to investigate random tensor networks numerically to check our scaling predictions. We note that although the critical bond dimension $D_c$ is non-universal and depends on the bulk lattice, we expect it to be low. In particular, it is possible that $1<D_c<2$, consistent with the natural expectation that random tensor networks with finite bond dimension are highly entangled. We expect the generalized PEPS wavefunctions~\eqref{eqWavefunction} to be useful to access a regime of non-integer bond dimension numerically.  

Another potentially interesting direction would be to incorporate the effects of global symmetries on the entanglement structure of RTN, which produce a local gauge structure in the bulk~\cite{2017arXiv171109941C}, and could potentially alter the universal scaling properties of the entanglement transition.

\label{SecDiscussion}

{\it Acknowledgments.}--- We thank E. Altman, J. Chalker, J. Erdmenger, M. Headrick, D. Huse, A. Nahum, H. Saleur, E. Tonni and X.-L. Qi for insightful discussions and useful comments. We are grateful to the KITP Program ``Synthetic Quantum Matter'', the KITP follow-on program ``Many-body localization'', the KITP program ``The Dynamics of Quantum Information'', as well as to the Aspen Workshop ``Entanglement Matters'', where parts of this work were carried out. This work is supported in part by the National Science Foundation  under Grants No. DMR-1653007 (ACP) and DMR-1309667 (AWWL). This work was supported by the US Department of Energy, Office of Science, Basic Energy Sciences, under Early Career Award No. DE-SC0019168 (RV). Part of this work was also performed at the Aspen Center for Physics, which is supported by National Science Foundation grant PHY-1066293, and at the KITP, which is supported by the National Science Foundation under Grants No. PHY11-25915 and NSF PHY-1748958.

\bibliography{References}
\bibliographystyle{apsrev4-1}

\end{document}